\documentclass[aps,prl,twocolumn,amsmath,amssymb,superscriptaddress,longbibliography]{revtex4-1}
\usepackage[english]{babel}
\usepackage{xcolor}
\usepackage{amssymb}
\usepackage{dcolumn}
\usepackage{bm}
\usepackage{graphicx}
\usepackage{amsmath}
\usepackage{braket}

\allowdisplaybreaks[4]   %Formula span

\usepackage{graphicx}        % standard LaTeX graphics tool

\graphicspath{{pict/}{}}

\usepackage[normalem]{ulem}

\usepackage{dcolumn}%%%%%new
\usepackage{bm}
\usepackage[pdfstartview=FitH, CJKbookmarks=true, bookmarksnumbered=true, bookmarksopen=true, colorlinks=true, pdfborder=001, citecolor=blue, linkcolor=blue, urlcolor=blue, linktocpage=true] {hyperref}

\begin{document}

\title{Self-Ordered Supersolid in Spinor Condensates with Cavity-Mediated Spin-Momentum-Mixing Interactions}

\author{Jingjun You}
\affiliation{Guangdong Provincial Key Laboratory of Quantum Metrology and Sensing $\&$ School of Physics and Astronomy, Sun Yat-Sen University (Zhuhai Campus), Zhuhai 519082, China}

\author{Su Yi}
\email{syi@itp.ac.cn}
\affiliation{CAS Key Laboratory of Theoretical Physics, Institute of Theoretical Physics, Chinese Academy of Sciences, Beijing 100190, China}
\affiliation{CAS Center for Excellence in Topological Quantum Computation, University of Chinese Academy of Sciences, Beijing 100049, China}
\affiliation{School of Physical Sciences, University of Chinese Academy of Sciences, Beijing 100049, China}

\author{Yuangang Deng}
\email{dengyg3@mail.sysu.edu.cn}
\affiliation{Guangdong Provincial Key Laboratory of Quantum Metrology and Sensing $\&$ School of Physics and Astronomy, Sun Yat-Sen University (Zhuhai Campus), Zhuhai 519082, China}

\date{\today}

\begin{abstract}  
Ultracold atoms with cavity-mediated long-range interactions offer a promising platform for investing novel quantum phenomena. Exploiting recent experimental advancements, we propose an experimental scheme to create self-ordered supersolid in spin-$1/2$ condensates confined within an optical cavity. The interplay of cavity and pump fields gives rise to supersolid square and plane wave phases, comprehensively described by the two-component Tavis-Cummings model. We show that the self-ordered supersolid phase exhibits an undamped gapless Goldstone mode over a wide parameter range. This proposal, achievable with current experimental setups utilizing identical laser configurations, is in contrast to the realization of checkerboard supersolidity, which hinges on constructing a $U(1)$ symmetry by utilizing two ${\cal Z}_2$ symmetries with precisely matched atom-cavity coupling in multimode resonators. By employing the superradiant photon-exchange process, we realize for the first time cavity-mediated spin-momentum-mixing interactions between highly correlated spin and momentum modes, analogous to that observed spin-mixing in spin-1 condensates. Our scheme provides a unique platform for realizing spin-momentum squeezing and spatially distributed multipartite entanglement. 
\end{abstract}

\maketitle

{\em Introduction}.---The experimental realization of ultracold atomic gases in optical cavities has unveiled unprecedented opportunities for simulating complex quantum matter~\cite{RevModPhys.85.553,Mivehvar:2021lpo,RevModPhys.95.035002,RevModPhys.85.299,RevModPhys.91.015006}, providing  powerful solvers for revealing fundamental physics~\cite{RevModPhys.73.565,RevModPhys.82.1041,RevModPhys.91.025005} and broad applications in quantum technology~\cite{RevModPhys.87.1379,RevModPhys.94.041003,RevModPhys.93.025005}. Among these advancements, the supersolid, which uniquely combines the dissipationless flow of superfluid and the long-range periodic density modulation of  crystalline order, stands as one of the most enigmatic quantum states of matter~\cite{RevModPhys.84.759}. The core challenge for supersolidity is requiring spontaneously breaking two mutually exclusive $U(1)$ symmetries, and its experimental exploration has undergone a tortuous development in solid helium~\cite{kim_probable_2004,balibar_enigma_2010,PhysRevLett.109.155301}, despite theoretical predictions dating back over half a century~\cite{PhysRevA.2.256,PhysRev.106.161}. Recently, supersolidity has been extensively studied both theoretically and experimentally in frustrated quantum magnets~\cite{PhysRevLett.131.116702}, high-pressure deuterium~\cite{PhysRevLett.128.045301}, and synthetic materials~\cite{PhysRevLett.130.226001,PhysRevLett.130.057001,PhysRevLett.127.247701}. 

The extraordinary capability of ultracold atoms, enriched with artificial gauge fields~\cite{NATYJ2011SOCAT,PRLZW2012SOCAT,SCPJW2016SOC,NSHL2016,PhysRevLett.130.156001,li2017stripe}, dynamical spin-orbital coupling (SOC)~\cite{PhysRevLett.112.143007,PhysRevA.89.011602,PhysRevLett.107.270401,PhysRevResearch.5.013002,PhysRevLett.123.160404}, and intrinsic dipolar interaction~\cite{PhysRevLett.128.195302,PhysRevLett.129.040403,guo2019low,PhysRevLett.131.223401}, provides new opportunities for exploration of supersolid phases and advanced many-body physics~\cite{PhysRevLett.128.103201,RevModPhys.80.885,RevModPhys.83.863,RevModPhys.83.1523,RevModPhys.91.015005,gross2017quantum}. In particular, the supersolid phase with a gapless Goldstone mode has been experimentally realized in spinless BEC coupled to two noninterfering standing-wave optical cavities~\cite{leonard_supersolid_2017,leonard_monitoring_2018} and two modes of the ring cavity~\cite{PhysRevLett.124.143602}. The mechanism for generating $U(1)$ symmetry employs by two ${\cal Z}_2$ symmetries with strictly equal couplings for two-mode cavities, analogous to constructing the $U(1)$ symmetric $XY$ model from two ${\cal Z}_2$ symmetric Ising models~\cite{leonard_supersolid_2017,leonard_monitoring_2018}. Recently, promising routes for the realization of supersolid phases by utilizing a confocal multimode cavity~\cite{PhysRevLett.131.173401,guo2021optical,PhysRevLett.120.123601,PhysRevLett.122.190801,PhysRevLett.124.033601} and axially elongated dipolar quantum droplets~\cite{tanzi2019supersolid,PhysRevLett.122.130405,PhysRevX.9.021012,PhysRevX.9.011051} are extensively explored. The unambiguous supersolidity for experimental observation and the corresponding enigmatic fundamental quantum properties are yet to be fully explored, despite rapid theoretical proposals and experimental advances ranging from condensed-matter physics to ultracold quantum gases and cavity-QEDs.

In this Letter, we propose a readily implementable experimental scheme to realize a supersolid square (SS) phase that features an undamped zero-energy Goldstone mode in spin-$1/2$ condensates coupled to an optical cavity. Unlike the experimentally observed Dicke superradiance with discrete ${\cal Z}_2$ symmetry breaking~\cite{baumann2010dicke,mottl2012roton,PhysRevLett.121.163601}, the self-ordered superradiant phase transition is fully characterized by two-component Tavis-Cummings model (TCM), indicative of broken continuous $U(1)$ symmetry. Notably, the laser configuration proposed here is akin to pioneering experiment for realizing dynamical SOC with respect to Dicke model~\cite{PhysRevLett.123.160404}. Although supersolidity has been achieved for spinless BEC coupled to multimode cavities~\cite{leonard_supersolid_2017,leonard_monitoring_2018,PhysRevLett.124.143602} needing strictly equal coupling strengths, our method only requires a simpler laser configuration and supersolid phase exists in a large parameter regime. Interestingly, our scheme also generates for the first time cavity-mediated spin-momentum-mixing interactions characterized by strong spin-momentum correlations, in which pairs of zero-momentum atoms spin flip to same internal state in different momentum modes that goes beyond weak spin-exchange collisions in spinor BEC~\cite{PhysRevLett.81.5257} and collective momentum-exchange processes in atom-cavity systems~\cite{luo2023cavity}. Our findings promise to significantly broaden the experimental scope for investigating multipartite entanglement between opposite atomic momenta~\cite{PhysRevLett.120.033601,PhysRevX.13.021031,lange2018entanglement,fadel2018spatial,kunkel2018spatially} and open up a new avenue for designing versatile quantum simulators within the realm of ultracold quantum gases~\cite{martin2013quantum,bohnet2016quantum,christakis2023probing,li2023tunable}. 

\maketitle
%\section{Model and Hamiltonian\label{Sec2}}

\begin{figure}
 \includegraphics[width=0.88\columnwidth]{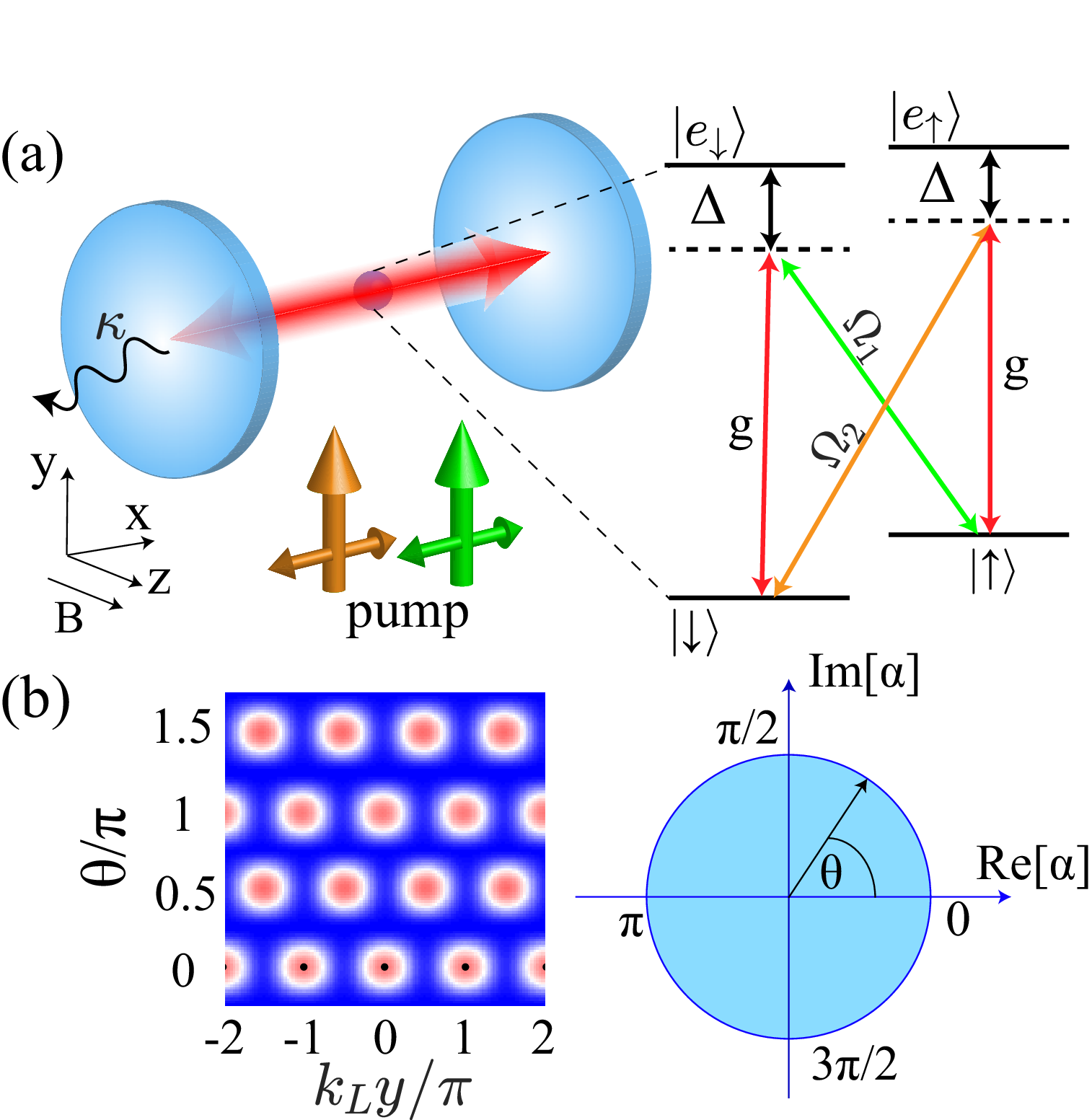}%
 \caption{(a) Schematic for creating two-component TCM. (b) Density distribution of self-ordered SS phase in terms of cavity amplitude ${\rm Re}[\alpha]$ and ${\rm Im}[\alpha]$ forms a circle. The profile exhibits a $\lambda/2$ period and corresponds to the positions changing continuously  along $y$-axis by varying ${\rm{arg}(\alpha)}$, demonstrating spontaneous continuous translational symmetry breaking. \label{model}
 }
\end{figure}

{\em Model}.---We consider an $N$ bosonic atomic $^{87}$Rb BEC consisting of two ground and two excited states trapped in a high-finesse optical cavity. Figure \ref{model}(a) illustrates the level diagram of atoms. An applied bias magnetic field ${\bf B}$ breaks the degeneracy of the ground (excited) state manifold  labeled as $|\uparrow\rangle$ and $|\downarrow\rangle$ ($|e_\uparrow\rangle$ and $|e_\downarrow\rangle$). The produced Zeeman shift between two hyperfine states $|\uparrow\rangle$ and $|\downarrow\rangle$ is given by $\hbar \omega_Z$ with their magnetic quantum numbers fulfilling $m_\uparrow = m_\downarrow  +1$ and $m_{e_\uparrow} = m_{e_\downarrow}  +1$. The transitions between $|\downarrow\rangle\leftrightarrow |e_\uparrow\rangle$ and $|\uparrow\rangle\leftrightarrow |e_\downarrow\rangle$ are driven by two transverse $\sigma$-polarized pump beams along $y$ axis, with Rabi frequencies $\Omega_{1,2}(y)=\Omega_{1,2}e^{ik_Ly}$. Here $k_L= 2\pi/\lambda$ is the wave vector of the laser field with $\lambda$ being the wavelength. The laser frequency difference is set to $2\omega_Z$ for compensating Zeeman shift. Dynamical SOC is crafted through the coupling of $|\sigma\rangle\leftrightarrow |e_\sigma\rangle$ ($\sigma=\uparrow$ and $\downarrow$) transitions by a $\pi$-polarized standing-wave cavity with the single atom-cavity coupling $g(x)=g\cos(k_L x)$. Collective Bragg scattering into $\sigma$-polarized cavity mode is mitigated under the conditions $|\omega_Z/g|\gg 1$~\cite{PhysRevLett.112.143007}.  

For large atom-pump detuning $|\Delta| \gg \{g,\Omega_{1,2}\}$, the atomic excited states $|e_{\uparrow,\downarrow}\rangle$ can be adiabatically eliminated and replaced by their steady-state solutions. Incorporating short-range collisional interactions, the effective many-body Hamiltonian of atom-cavity reads 
\begin{align}
\hat {\cal H}_0=&\hbar\Delta_c\hat{a}^{\dag}\hat{a} + \sum_{\sigma\sigma'}\int d{\mathbf r}\hat\psi_{\sigma}^{\dag}({\mathbf r})[\hat h_{\sigma\sigma'}+V_{b}({\mathbf r})\delta_{\sigma\sigma'}]\hat\psi_{\sigma'}({\mathbf r})  \nonumber \\
+&\frac{1}{2}\sum_{\sigma \sigma'}\frac{4\pi\hbar^2 a_{\sigma\sigma'}}{m}\int d{\bf r}\hat{\psi}^\dag_{\sigma}({\bf r})\hat{\psi}^\dag_{\sigma'}({\bf r})\hat{\psi}_{\sigma'}({\bf r})\hat{\psi}_{\sigma}({\bf r}),\label{manyh}
\end{align}
where $m$ is the mass of atom, $\hat {a}$ is the annihilation operator of cavity, and $\hat\psi_{\sigma}$ represents the annihilation bosonic field operator for spin-$\sigma$ atom.  $\Delta_{c}$ is the pump-cavity detuning, $V_b({\mathbf r})$ is the external uniform potential of box trap, and $a_{\sigma\sigma'}$ are $s$-wave scattering lengths for intraspecies ($\sigma=\sigma'$) and interspecies ($\sigma\neq\sigma'$) spin atoms. Additionally, the single-particle Hamiltonian satisfies
\begin{align}
    \hat{\boldsymbol h}=\frac{{\mathbf p}^2}{2m}\hat{I} + \hbar\left(
        \begin{array}{cc}
            \hat{M}_0(x)+\delta/2& \hat{M}_-(x,y)\\
            \hat{M}_-^\dagger(x,y)&\hat{M}_0(x)-\delta/2
        \end{array}
    \right),
\end{align}
where $\delta$ is the tunable two-photon detuning, $\hat{M}_-(x,y)=(g_1 \hat{a}e^{ik_Ly} + g_2 \hat{a}^\dagger e^{-ik_Ly}) \cos(k_Lx)$ is cavity-mediated Raman coupling with $g_{1,2} = -g\Omega_{1,2}/\Delta$ being the maximum scattering rate, and $\hat{M}_0(x)= -U_0 \cos^2(k_L x) \hat{a}^\dagger \hat{a}$ is optical lattice with Stark shift $U_{0}=g^2/\Delta$.  The spin-flip interaction $\hat{M}_-$ integrates two dynamical SOC~\cite{PhysRevLett.112.143007,PhysRevA.89.011602}, emerging from the interference between cavity and two classical pump fields. This interplay of dual dynamical SOC fosters spatially translational invariant crystalline orders within the condensates wave functions. 

To advance, cavity field $\hat{a}$ is adiabatically eliminated in the far dispersive limit $|\Delta_c/g_{1,2}| \gg 1$ since its dynamical evolution is much faster than external atomic motion. The steady-state intracavity amplitude satisfies 
\begin{align}
\alpha = \langle\hat{a}\rangle =({ g_1 {{\Xi}^\dagger_1} + g_2 {{\Xi}_2}})/({-\tilde{\Delta}_c + i\kappa}), \label{cavity}
\end{align}
where $\kappa$ is the cavity decay, $\tilde{\Delta}_c =(\Delta_c-U_0{\cal B})$ represents the $N$-dependent dispersive shift of cavity with ${\cal B}=\sum_\sigma\langle\psi_\sigma |\cos^2(k_Lx)|\psi_\sigma \rangle$, $\Xi_1 = \langle\psi_\uparrow |\cos(k_Lx)e^{ik_Ly}|\psi_\downarrow \rangle/N$ and $\Xi_2 = \langle\psi_\uparrow |\cos(k_Lx) e^{-ik_Ly}|\psi_\downarrow \rangle/N$ are order parameters dictating the configurations of ground state condensates wave functions. Consequently, the self-ordered superradiance corresponds to a finite $\alpha$ due to $\Xi_{1,2} \neq 0$ for plane wave (PW) and SS phases. We neglect the atom-cavity entanglement in Eq.~(\ref{cavity}) for considering a moderate photon emissions with $|\alpha|\sim 1$. The entanglement between photon and condensates, along with the quantum noise of cavity can be ignored when $|\tilde{\Delta}_c/\kappa|\gg 1$~\cite{nagy2008self,maschler2008ultracold}. 

\begin{figure}
 \includegraphics[width=0.86\columnwidth]{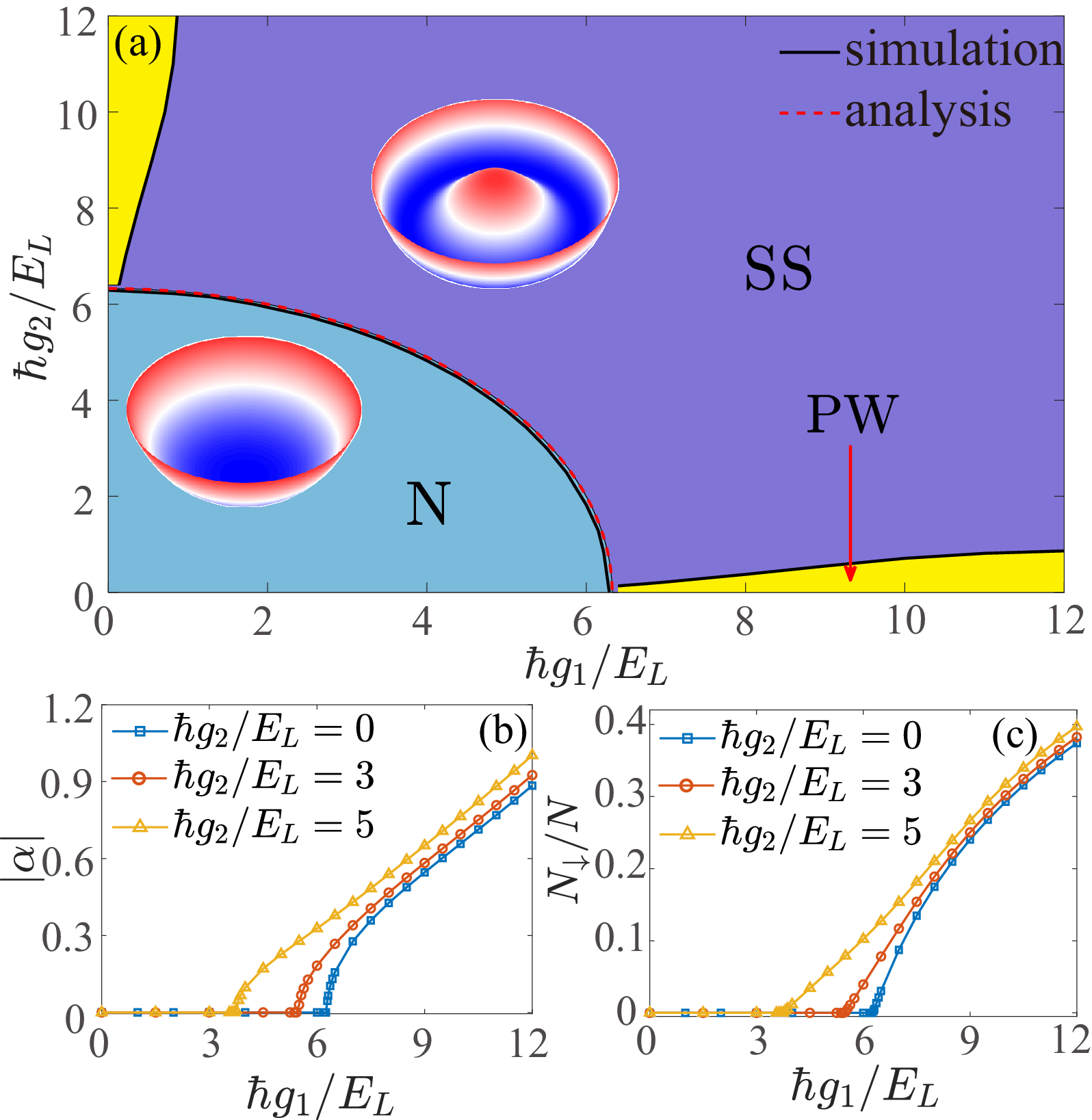}%
 \caption{(a) Phase diagram of ground state on $g_1$-$g_2$ parameter plane. The solid (dashed) line  marks numerical (analytical) phase boundaries. (b) and (c) show, respectively, $g_1$ dependence of $|\alpha|$ and $N_{\downarrow}$ for different values of $g_2$. \label{pd}}
\end{figure}

{\em Spin momentum mixing interactions}.---To deeper understanding of underlying physics, the self-organized superradiance can be comprehensively characterized by the two-component TCM Hamiltonian ($\hbar=1$)~\cite{SM}
\begin{align}\label{twoTCM}
   \hat{\mathcal{H}_1}& =\! \tilde{\Delta}_c \hat{a}^{\dagger} \hat{a}  +\! \omega_0\hat{J}_z +\! \frac{g_1}{\sqrt{2}}(\hat{a}\hat{J}_-^{(1)} \!+\! \frac{g_2}{g_1}\hat{a}^\dagger\hat{J}_-^{(2)} \!+\! {\rm H.c.}),
\end{align}
which describes two-component two-level bosonic atoms coupled to a quantized cavity.  $\omega_0= 2E_L/\hbar-\delta$ is the effective detuning of atomic field,  $\hat{J}_{-}^{(1)} = \hat b_{\uparrow,0}^\dag\hat b_{\downarrow,1}$, $\hat{J}_{-}^{(2)} = \hat b_{\uparrow,0}^\dag\hat b_{\downarrow,2}$, and $\hat{J}_{z} = ( \hat b_{\downarrow,1}^\dag\hat b_{\downarrow,1} +\hat b_{\downarrow,2}^\dag\hat b_{\downarrow,2}- \hat b_{\uparrow,0}^\dag\hat b_{\uparrow,0})/2$ are collective spin operators. Here $\hat b_{\uparrow,0}$ and $\hat b_{\downarrow,1}$ ($\hat b_{\downarrow,2}$) represent the annihilation bosonic operators for zero- and nonzero-momentum with $|k_{x},k_{y}\rangle=|\pm k_L, k_L\rangle$ ($|\pm k_L, -k_L\rangle$) in $|\uparrow\rangle$ and $|\downarrow\rangle$ atoms. Accordingly, the critical Raman coupling for two-component TCM satisfies
\begin{align}
 [g_1^2+g_2^2]_{\rm {cr}}=2(\tilde{\Delta}^2+\kappa^2)^{1/2}\omega_0/N. \label{cr}
\end{align}
Interestingly, the threshold of superradiant phase transition is combination of two cavity-mediated Raman couplings, revealing a high ground-state degeneracy. 

Unlike dynamical SOC for Dicke superradiance~\cite{PhysRevLett.123.160404}, the $U(1)$-symmetric $\hat{\mathcal{H}_1}$ is gauge invariant under the unitary transformation ${\mathcal R}_{\theta}^\dagger(\hat{a},\hat{J}_-^{(1)},\hat{J}_-^{(2)}){\mathcal R}_{\theta} =(\hat{a} e^{i\theta},\hat{J}_-^{(1)}e^{-i\theta},\hat{J}_-^{(2)}e^{i\theta})$ with respect to the operator $\mathcal{R}_\theta = \exp[i\theta(\hat{a}^\dagger\hat{a}- \hat{b}^\dagger_{\downarrow,1}\hat{b}_{\downarrow,1}+\hat{b}^\dagger_{\downarrow,2}\hat{b}_{\downarrow,2})]$~\cite{PhysRevE.67.066203,PhysRevLett.112.173601}. Indeed, $\hat{\mathcal{H}_1}$ is realized within the single recoil scattering limit~\cite{baumann2010dicke}, excluding contact interactions as well. The microscopic picture of superradiance coherently transfers the atomic motional ground state $|k_x, k_y\rangle=|0,0\rangle$ to the equal-energy excited momentum states $|k_{x},k_{y}\rangle=|\pm k_L,\pm k_L\rangle$ due to the cavity-mediated dynamical SOC mechanisms. We check that the threshold of superradiance is robust against the slightly variations of $s$-wave scattering length.

To characterize new properties of superradiance, one can derive cavity-mediated collective interactions of atomic fields by integrating out the cavity field~\cite{SM}
\begin{align}\label{manyint}
 \hat{\mathcal{H}}_{2}&=\frac{1}{2}\int d\mathbf{r}d\mathbf{r^\prime} U_{\rm eff}(\mathbf{r},\mathbf{r^\prime}) \{V_1\hat{S}_-(\mathbf{r})\hat{S}_+(\mathbf{r^\prime})  \nonumber \\
 & + V_2\hat{S}_+(\mathbf{r})\hat{S}_-(\mathbf{r^\prime}) + V_3[\hat{S}_+(\mathbf{r})\hat{S}_+(\mathbf{r^\prime})+ {\rm H.c.}] \},
\end{align}
where $U_{\rm eff}(\mathbf{r},\mathbf{r^\prime})= 2\cos(k_Lx)\cos(k_Lx^\prime)e^{ik_L(y-y^\prime)}$
is the long-range  potential, $\hat{S}_+(\mathbf{r})=\hat{\psi}_\uparrow^\dagger(\mathbf{r})\hat{\psi}_\downarrow(\mathbf{r})$ is the spin operator,  $V_{j=1,2} =-{\tilde{\Delta}_cg_j^2}/({\tilde{\Delta}_c^2 + \kappa^2})$ and $V_3= -{\tilde{\Delta}_cg_1g_2}/({\tilde{\Delta}_c^2 + \kappa^2})$ are tunable strengths of two-body interactions. Remarkably, the first two terms of $\hat{\mathcal{H}}_{2}$ represent the long-range spin-exchange interaction between spin-$\uparrow$ and $\downarrow$ atoms, which has been extensively studied including atom-cavity superradiance~\cite{norcia2018cavity,muniz2020exploring}. The third term introduces a two-axis twisting interaction via a superradiant photon-exchange process originally proposed by Kitagawa and Ueda~\cite{PhysRevA.47.5138}, which does not conserve the atom number in the individual spin state~\cite{chomaz2022dipolar,ma2011quantum,PhysRevA.47.5138}.  It is worth noting that these interactions  provide a new avenue for exploring self-organized superradiance in spinor condensates.

By performing a few-mode expansion~\cite{SM}, the spin-momentum-mixing interactions in terms of spin and momentum modes are given by
\begin{align}\label{orbitH}
\hat{\mathcal{H}}_{3}&=\sum_{j=1,2}\frac{V_j}{2}\hat{b}_{\uparrow,0}^\dagger\hat{b}_{\downarrow,j}^\dagger\hat{b}_{\uparrow,0}\hat{b}_{\downarrow,j} +\!\frac{V_3}{2}(\hat{b}_{\uparrow,0}^\dagger\hat{b}_{\uparrow,0}^\dagger\hat{b}_{\downarrow,1}\hat{b}_{\downarrow,2}+\!{\rm H.c.}).
\end{align} 
The first term denotes collective momentum-exchange interactions~\cite{luo2023cavity}, referred to as one-axis twisting. Notably, the second term originating from two-axis twisting provides a new source for the deterministic generation of entangled momentum-correlated pairs from zero-momentum condensates. This controllable spin-momentum-mixing interactions is similar to spin-mixing in spin-1 BEC under the single-mode approximation~\cite{PhysRevLett.81.5257}. We should note that Hamiltonian (\ref{orbitH}) will dominate self-organized condensates wave functions of ground states subsequent to superradiance, significantly surpass the typical two-body collisional interaction. Remarkably, the magnitude of  $N{V}_3$ is typically on the order of tens of kilohertz, vastly exceeding the spin-mixing rate (a few tens of hertz) achievable in experimental engineering twin-Fock states for $^{87}$Rb condensates~\cite{luo2017deterministic}. Compared to the experimentally realized momentum-exchange interactions~\cite{luo2023cavity}, Eq.~(\ref{orbitH}) undergoing cavity-mediated spin-momentum-mixing dynamics is deeply entangled, enabling the creation of novel nonclassical states highly correlated in different momentum modes, e.g., spatially separated deterministic entanglement~\cite{PhysRevLett.120.033601,PhysRevX.13.021031,lange2018entanglement,fadel2018spatial,kunkel2018spatially}.

To gain more insight, Hamiltonian $\hat{\mathcal{H}}_{3}$ can be simplified to a one-axis twisting model in well-defined orbital angular momentum operators
\begin{align}\label{orbital}
  \hat{\mathcal{H}}_{\rm eff}&=\frac{1}{4}V_3\hat{L}_+\hat{L}_-=\frac{1}{4}V_3(\hat{L}^2 -\hat{L}_z^2+ \hat{L}_z),
\end{align}
when $g_1=g_2$. The raising and lowering operators $\hat{L}_+=\hat{L}_-^\dagger=\sqrt{2}(\hat{b}^\dagger_{\downarrow,2}\hat{b}_{\uparrow,0}+\hat{b}^\dagger_{\uparrow,0}\hat{b}_{\downarrow,1})$ obey the angular momentum commutation relations $[\hat{L}_+,\hat{L}_-]=2\hat{L}_z$ and $[\hat{L}_z,\hat{L}_\pm]=\pm \hat{L}_\pm$ with $\hat{L}_z=\hat{b}_{\downarrow,1}^\dagger\hat{b}_{\downarrow,1}-\hat{b}_{\downarrow,2}^\dagger\hat{b}_{\downarrow,2}$. We should emphasize that the operator $\hat{L}$ acts on the coupled spin-1 spin-momentum degree of freedom in contrast to the spinor condensates.

%\section{Results\label{Sec3}}
{\em Self-ordered phase diagram}.---We study ground-state phases by solving Gross-Pitaevskii equations via imaginary time evolution. In mean-field framework, atomic and cavity field operators are replaced by $\psi_\sigma=\langle \psi_\sigma\rangle$ and $\alpha=\langle \hat{a}\rangle$ and calculated in a self-consistent manner. We consider $N = 10^5$ BEC initially in $|\uparrow\rangle$ state and confined in a circular optical box trap~\cite{Nature2016,PhysRevLett.110.200406}. Unlike pancake-shaped harmonic trap, the elaborately selected uniform potential preserves a continuous translational symmetry. Specifically, we take single-photon recoil energy $E_L/\hbar = 3.53{\rm kHz}(2\pi)$ with the wavelength $\lambda = 2\pi/k_L= 803.2 {\rm nm}$, cavity decay rate $\kappa= 100E_L/\hbar$, two-photon detuning $\delta = -2E_L/\hbar$, Stark shift $U_0=10E_L/\hbar$, and pump-cavity detuning $\tilde{\Delta}_c= U_0N/2$. The $s$-wave scattering lengths for collisional interactions are $a_{\downarrow\downarrow}=a_{\uparrow\downarrow}\approx a_{\uparrow\uparrow}=50a_B$ with $a_B$ being the Bohr radius. We emphasize that the threshold of superradiance is largely unaffected by contact interaction, attributed to the significant cavity-mediated collective interactions. To this end, the free controllable parameters are Raman couplings $g_{1}$ and $g_{2}$. 

Figure \ref{pd}(a) summarized the ground-state phase diagram. The interplay of Raman couplings $g_1$ and $g_2$ leads to two self-ordered superradiant PW and SS phases with nonzero intracavity amplitude $\alpha\neq0$. In the absence of superradiance, the superfluid phase is denoted by "N". It is apparent that SS and PW phases appear symmetrically in phase diagram, highlighting that SS phase does not require $g_1/g_2=1$. This observation marks a departure from checkerboard supersolidity, which demands strictly equal couplings to maintain $U(1)$ symmetry in dual-mode cavities~\cite{leonard_supersolid_2017,leonard_monitoring_2018}. For a relatively small $g_1$ ($g_2$), the system enters PW phase without a crystalline order induced by dynamical SOC~\cite{PhysRevResearch.5.013002,PhysRevLett.123.160404}.

The transition from N to self-ordered SS phase is attributed to superradiance in two-component TCM, corresponding to the effective potential of ground states extending from a minimum in the origin to a sombrero shape with a circular valley of degenerate minima. The observed quarter circle on $g_1$-$g_2$ plotted in Fig.~\ref{pd}(a) further reveals the high ground-state degeneracy. The analytical threshold for superradiance (red dashed line) is in strong agreement with numerical simulations (black solid line), incorporating atomic collision and cavity dispersion. In Fig.~\ref{pd}(b) and \ref{pd}(c), we show cavity amplitude ($|\alpha|$) and spin-$\downarrow$ population ($N_\downarrow$) versus $g_1$ for various $g_2$ values, serving as order parameters to distinguish between N and superradiant phases. This marks the breaking of $U(1)$ symmetry from vacuum to a finite value spontaneously. Additionally, a large $g_2$ lowers the $g_1$ threshold for superradiance, as predicted  analytically in Eq.~(\ref{cr}). 
  
\begin{figure}
 \includegraphics[width=0.85\columnwidth]{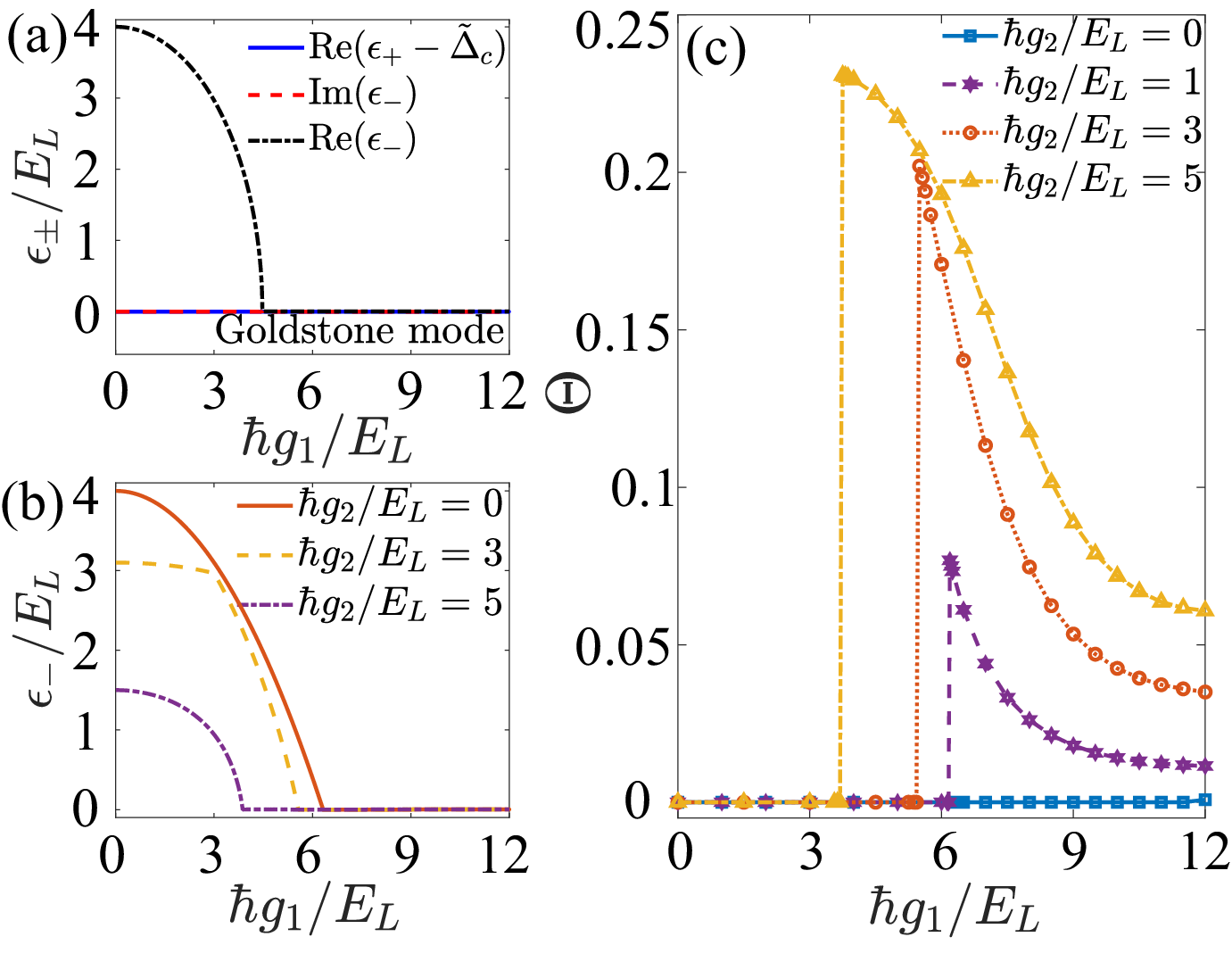}%
  \caption{(a) $g_1$ dependence of collective excitations $\epsilon_{\pm}$ with $g_1/g_2=1$. (b) Lower branch $\epsilon_{-}$ and (c) order parameter $\Theta$ as a function of $g_1$ for different values of $g_2$. \label{golden}}
\end{figure}

To demonstrate the rigidity of self-ordered superradiant phases, we calculate the collective excitation of two-component TCM. By diagonalizing the Hopfield-Bogoliubov matrix~\cite{SM}, a gapless Goldstone mode of low-energy excitation is identified with respect to superradiance of spontaneous $U(1)$ symmetry broken, as displayed in Fig.~\ref{golden}(a). Despite the nonzero cavity dissipation included, the lifetime of this zero-energy mode is far exceeding that of quantized cavity and roughly undamped with ${\rm Im}(\epsilon_{-})/\kappa\sim 10^{-4}$. More importantly, this zero-energy Goldstone mode persists in superradiant phases whether $g_1$ equals $g_2$ or not [Fig.~\ref{golden}(b)], which will significantly enhance the experimental feasibility in atom-cavity QEDs~\cite{leonard_supersolid_2017,leonard_monitoring_2018,PhysRevLett.124.143602}.

The order parameter of $\alpha$ characterizes superradiant phase transition from vacuum ($\alpha=0$) to a finite value ($\alpha\neq 0$), but it cannot discern the transition from PW to SS phase. To address this, we introduce an order parameter $\Theta \equiv \langle \psi_\downarrow | \cos(2k_L x) \cos(2k_L y)|\psi_\downarrow\rangle /N$, which quantifies the periodic density modulation via the configuration of the condensates. In Fig.~\ref{golden}(c), we depict $g_1$ dependence of $\Theta$ for different
values of $g_2$. Compared to PW phase with $\Theta\sim0$, SS phase shows a large value of $\Theta$, revealing a strong periodic density modulation of ground state. Particularly for $g_1/g_2\neq1$, the emergence of a self-ordered crystalline structure with respect to a zero-energy Goldstone mode is also confirmed. The large $g_2$ corresponds to a heightened $\Theta$, signifying a strong self-ordered crystalline arrangement. Moreover, $\Theta$ exhibits a sharp increase from N to SS phases, facilitating the natural monitoring the superradiant phase transition.

 \begin{figure}
 \includegraphics[width=0.86\columnwidth]{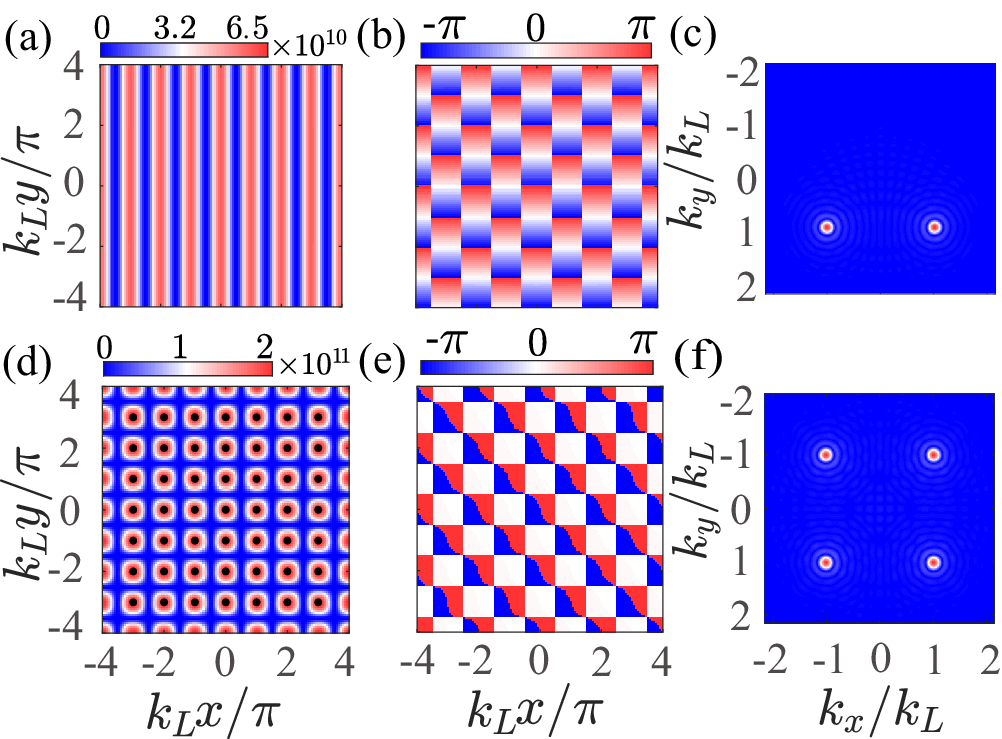}%
  \caption{(Color online). The condensates wave functions of PW (row 1) and SS phase (row 2) for $(g_1, g_2) =(0, 7)E_L/\hbar$ and $(7, 7)E_L/\hbar$ in spin-$\downarrow$ atom, respectively. Column 1-3 show the density $\rho_\downarrow$ (units of ${\rm cm}^{-2}$), relative phase $\Delta \phi$, and momentum distribution.  \label{pm}}
\end{figure}

Figure \ref{pm} shows typical condensates wave functions for two superradiant phases. We find that the self-ordered crystalline order always appears in less populated spin-$\downarrow$ state since $\uparrow$-component is populated dominantly in numerical simulations. The PW phase presents a nearly structureless density profile with staggered $\lambda$-periodic phase modulation along $y$ axis and $\lambda/2$-periodic density stripe along $x$ axis, created by the cavity-induced optical lattice. For minimizing energy, the relative phase of condensates wave functions requires $\Delta \phi=\arg(\psi_\downarrow)-\arg(\psi_\uparrow)= k_Ly+\pi$ ($k_Ly$) when $\cos(k_Lx)=1$ ($-1$)~\cite{PhysRevLett.112.143007}. With this correspondence, the momentum-space distribution exhibits two peaks at $|\pm k_L, -k_L\rangle$, as shown in Fig.~\ref{pm}(c). Furthermore, the weakly 2D periodic density modulation is in good agreement with $\Theta\sim0$. Similar to TCM superradiance~\cite{PhysRevResearch.5.013002}, the observed PW phase resulted from dynamical SOC is dominated by individual $g_2$ ($g_1$) Raman coupling with $g_2/g_1\gg 1$($g_1/g_2\gg 1$) in Fig.~\ref{pd}(a). Compared to experimentally observed PW phase with dynamical SOC and a gapped roton mode~\cite{PhysRevLett.123.160404}, our finding demonstrates the spontaneous $U(1)$ symmetry breaking and gapless Goldstone mode excitation [Fig.~\ref{golden}(b)].
 
When $g_1$ and $g_2$ are comparable, the system transitions into an interesting SS phase characterized by significant periodic density modulation with a high $\Theta$ value, distinguishing it from the PW phase. Notably, the self-ordered density profile depends on cavity phase angle ${\rm arg}(\alpha)$, corresponding to the positions changing continuously along $y$ axis [Fig.~\ref{model}(b)], which is unambiguous demonstration of supersolidity combined with rigidity of the gapless Goldstone mode. This phenomenon of self-organization in superfluid quantum gases that cavity amplitude $\alpha$ is self-consistently determined by the atomic wave function~\cite{baumann2010dicke}, originates from the broken continuous translational symmetry inherent in two-component TCM superradiance. For ${\rm arg}(\alpha)=0$, the sites of peak density satisfy $\cos(k_Ly)\cos(k_Lx)=+1(-1)$ with a corresponding relative phase $\Delta \phi=\pi(0)$. This setup yields atomic momentum distribution at $(k_x, k_y)=(\pm k_L, \pm k_L)$ and matches the configurations of crystalline square lattice pattern.

For experimental observation, the  transition into superradiant phases, characterized by a nonzero photon number can be monitored by measuring the inherent leakage of cavity~\cite{baumann2010dicke}. Distinguishing between PW and SS phases can be achieved by detecting atomic momentum distribution via spin-sensitive absorption images~\cite{PhysRevLett.121.163601} or using Bragg scattering techniques~\cite{li2017stripe}. Additionally, the presence of the zero-energy Goldstone mode insensitive to cavity dissipation can be identified through spectroscopic measurement for ultracold quantum gases~\cite{leonard_monitoring_2018}. Notably, the SS phase exhibiting the spatial periodicity of $\lambda/2$ is different from the previously realized $\lambda$ periodic ${\cal Z}_2$-broken checkerboard lattice supersolid involving both scalar and spinor condensates within a cavity~\cite{baumann2010dicke,mottl2012roton,PhysRevLett.121.163601}. 

%\section{Conclusion\label{Sec4}}
{\em Conclusion}.---We present an experimental scheme to generate self-ordered supersolid for cavity-coupled spinor condensates. The supersolid square phase with combination of continuous translational symmetry breaking and the undamped gapless Goldstone mode, is characterized by two-component TCM superradiance. Our proposal also highlights the enhanced experimental feasibility compared to pioneer studies and can be readily tested in realizable experiments. In contrast to the realized dynamical SOC with discrete ${\cal Z}_2$ symmetry in Benjamin L. Lev group~\cite{PhysRevLett.123.160404}, our work preserves continuous translational symmetry using the same laser configuration. Remarkably, the mechanism for generating supersolidity does not require equal couplings for the two cavity-mediated Raman processes, significantly simplifying experimental feasibility in cavity QEDs~\cite{leonard_supersolid_2017,leonard_monitoring_2018,PhysRevLett.124.143602,PhysRevLett.131.173401,guo2021optical,PhysRevLett.120.123601,PhysRevLett.122.190801,PhysRevLett.124.033601}. Furthermore, the cavity-mediated spin-momentum-exchange interactions, which intricately correlates spin and momentum modes, are realized for the first time. Compared to weak spin-mixing in spin-1 condensates and collective momentum-exchange interactions, the emergence of cavity-mediated spin-momentum-mixing interactions may facilitate the exploration of entanglement-enhanced metrology~\cite{luo2017deterministic}, spin-momentum squeezing~\cite{PhysRevA.106.043711,finger2023spin}, and spatially separated multipartite entanglement using highly correlated momentum modes~\cite{PhysRevLett.120.033601,PhysRevX.13.021031,lange2018entanglement,fadel2018spatial,kunkel2018spatially}.

\begin{acknowledgments}
{This work was supported by the NSFC (Grants No. 12274473 and No. 12135018), by the National Key Research and Development Program of China (Grant No. 2021YFA0718304), and by the Strategic Priority Research Program of CAS (Grant No. XDB28000000).}
\end{acknowledgments}

\newpage{}
\appendix   %�
\setcounter{table}{0}   %从0开始编号，显示出来表会A1开始编号
\setcounter{figure}{0}
\setcounter{equation}{0}
%定义编号格式，在数字序号前加字符“A"
% \renewcommand{\thetable}{A\arabic{table}}
\renewcommand{\thefigure}{A\arabic{figure}}
\section{Atom-cavity Hamiltonian}
In this section, we present the derivation of the effective atom-cavity Hamiltonian for the displayed laser configuration and level diagram in Fig.\ref{model} of the main text. We firstly ignores the two-body $s$-wave collisional interaction. Under the rotating-wave approximation, the single-particle Hamiltonian of the atom-cavity system excluding the kinetic energy is given by 
 \begin{align}\label{begin}
  \hat {\boldsymbol h}_0/\hbar &= \Delta_c \hat{a}^\dagger\hat{a} + \frac{\delta}{2}(\hat{c}^\dagger_\uparrow\hat{c}_\uparrow-\hat{c}^\dagger_\downarrow\hat{c}_\downarrow)+ \Delta_1\hat{e}^\dagger_\downarrow\hat{e}_\downarrow + \Delta_2\hat{e}^\dagger_\uparrow\hat{e}_\uparrow  \nonumber \\
&+ g \cos(k_Lx)(\hat{c}^\dagger_\downarrow\hat{a}^\dagger\hat{e}_\downarrow + \hat{c}^\dagger_\uparrow\hat{a}^\dagger\hat{e}_\uparrow +{\rm H.c.}) \nonumber \\ 
    &+ (\Omega_1\hat{e}^\dagger_\downarrow\hat{c}_\uparrow e^{-ik_Ly} + \Omega_2\hat{e}^\dagger_\uparrow\hat{c}_\downarrow e^{-ik_Ly} + {\rm H.c.}),
 \end{align} 
 where $\hat{a}$ is the annihilation operator of the optical cavity and $\hat{c}_\sigma$ ($\hat{e}_\sigma$) is the annihilation operator of the atomic field of the corresponding ground (excited) state. $\Delta_c$ is the pump-cavity detuning, $\Delta_{1,2}$ is the atom-pump detuning, $\delta$ is the tunable two-photon detuning, and $g(x)=g\cos(k_Lx)$ is the single atom-cavity coupling. 

Taking into account the atomic spontaneous emissions of excited states ($\gamma_i$) and cavity decay ($\kappa$) for completeness, the Heisenberg equations of motion for cavity and atomic field operators are given by 
\begin{align}
    i \dot{\hat{c}}_\downarrow &= -\delta\hat{c}_\downarrow/2 + \Omega_2\hat{e}_\uparrow e^{ik_Ly} +g\cos(k_Lx)\hat{a}^\dagger\hat{e}_\downarrow \nonumber \\
    i \dot{\hat{c}}_\uparrow &= \delta\hat{c}_\uparrow/2 + \Omega_1\hat{e}_\downarrow e^{ik_Ly} +g\cos(k_Lx)\hat{a}^\dagger\hat{e}_\uparrow \nonumber \\
    i \dot{\hat{e}}_\downarrow &= (\Delta_1-i\gamma_1)\hat{e}_\downarrow +\Omega_1 \hat{c}_\uparrow e^{-ik_Ly} +g\cos(k_Lx)\hat{a}\hat{c}_\downarrow    \nonumber \\
    i \dot{\hat{e}}_\uparrow &= (\Delta_2-i\gamma_2)\hat{e}_\uparrow +\Omega_2 \hat{c}_\downarrow e^{-ik_Ly} +g\cos(k_Lx)\hat{a}\hat{c}_\uparrow   \nonumber \\
    i \dot{\hat{a}} &= (\Delta_c-i\kappa)\hat{a} + g\cos(k_Lx)(\hat{c}_\downarrow^\dagger\hat{e}_\downarrow + \hat{c}_\uparrow^\dagger\hat{e}_\uparrow).  
\end{align}
In  the large atom-pump detuning limit $|\Delta_i|\gg\{g,\Omega_{1,2},\kappa,\gamma_i\}$ the dynamics of excited states reach equilibrium state faster than the ground states and photon field, which allows one to adiabatically eliminate the electronically excited states by making $i\dot{\hat{e}}_\sigma = 0$, which yields
\begin{align}\label{ade}
    \hat{e}_\downarrow&\approx -1/\Delta_1(\Omega_1 \hat{c}_\uparrow e^{-ik_Ly} +g\cos(k_Lx)\hat{a}\hat{c}_\downarrow)  \nonumber \\
    \hat{e}_\uparrow&\approx -1/\Delta_2(\Omega_2 \hat{c}_\downarrow e^{-ik_Ly} +g\cos(k_Lx)\hat{a}\hat{c}_\uparrow). 
\end{align}

Considering $\Delta_1\approx \Delta_2=\Delta$, the resulting single-particle Hamiltonian is given as
\begin{widetext}
\begin{eqnarray}
    \hat {\boldsymbol h} =\frac{{\mathbf p}^{2}}{2m}\hat I +
    \hbar\left(\!
    \begin{array}{cc}
        -U_0 \hat{a}^\dagger \hat{a} \cos^2(k_L x) +\delta/2&   (g_1 \hat{a}e^{ik_Ly} + g_2 \hat{a}^\dagger e^{-ik_Ly}) \cos(k_Lx)  \\
     (g_1 \hat{a}^\dagger e^{-ik_Ly} + g_2 \hat{a} e^{ik_Ly})\cos(k_Lx) & -U_0 \hat{a}^\dagger \hat{a} \cos^2(k_L x) -\delta/2
    \end{array}
    \!\right),\!\;
\end{eqnarray}
\end{widetext}
where $g_{1,2} = -g\Omega_{1,2}/\Delta$ corresponds to the maximum scattering rate and $U_{0}=g^2/\Delta$ is the Stark shift. The effective many-body interaction Hamiltonian of atom-cavity can be written as
\begin{align}
{\cal \hat{H}}_0 &= {\cal \hat{H}}_c + {\cal \hat{H}}_{a} + {\cal \hat{H}}_{\rm ac},  \label{smmany}
\end{align}%
where
\begin{align}
    {\cal \hat{H}}_c &= \hbar(\Delta_c-i\kappa)\hat{a}^{\dag}\hat{a}, \nonumber \\
    {\cal \hat{H}}_{\rm a} &= \int d {\mathbf r} \{ [\hat{M}_{0}(x) + V_{b}({\mathbf r})]\sum_{\sigma}\hat{\psi}_\sigma^\dag({\mathbf r})\hat{\psi}_\sigma ({\mathbf r}) \nonumber \\
    &+\frac{\delta}{2}[\hat{\psi}_\uparrow^\dag({\mathbf r})\hat{\psi}_\uparrow ({\mathbf r}) - \hat{\psi}_\downarrow^\dag({\mathbf r})\hat{\psi}_\downarrow({\mathbf r}) ]\} \nonumber \\
    &+\sum_{\sigma \sigma'}g_{\sigma\sigma'}\int d{\bf r}\hat{\psi}^\dag_{\sigma}({\bf r})\hat{\psi}^\dag_{\sigma'}({\bf r})\hat{\psi}_{\sigma'}({\bf r})\hat{\psi}_{\sigma}({\bf r}), \nonumber \\
    {\cal \hat{H}}_{\rm ac} &= \int d {\mathbf r} \{ (g_1 \hat{a}e^{ik_Ly} + g_2 \hat{a}^\dagger e^{-ik_Ly}) \nonumber \\
    &\times\cos(k_Lx)  \hat{\psi}_\uparrow^\dag({\mathbf r})\hat{\psi}_\downarrow({\mathbf r}) +  \rm{H.c.}\}.  \nonumber 
\end{align}
Here $\hat{\psi}_{\sigma}({\mathbf r})$ denotes the annihilation bosonic operator for atomic field of spin-$\sigma$ atom. ${\cal \hat{H}}_c$ represents the Hamiltonian of optical cavity with nonzero cavity dissipation. ${\cal \hat{H}}_a$ denotes the atomic Hamiltonian for pseudospin-$1/2$ system, including the optical lattice with $\hat{M}_{0}(x)=-U_0 \hat{a}^\dagger \hat{a} \cos^2(k_L x)$, external trapping potential $V_b({\mathbf r})$, effective Zeeman field, and two-body collisional interaction strength $g_{\sigma\sigma^\prime}=4\pi\hbar^2a_{\sigma\sigma^\prime}/m$. The last term ${\cal \hat{H}}_{\rm ac}$ denotes the cavity-condensates interaction originating from the photon superradiance due to combination of two pump fields and one quantized single-mode cavity.

Then the  dynamical equations for the atom and cavity operators of the pseudospin-$1/2$ system read
\begin{align}
    i\dot{\hat{a}} & = (\tilde{\Delta}_c -
    i\kappa)\hat{a} + g_1 {\hat{\Xi}_1}^\dagger+ g_2 {\hat{\Xi}_2},\nonumber \\
    i\dot{\hat{\psi}}_\uparrow & = [-U_0\hat{a}^\dagger \hat{a}\cos^2(k_Lx) + V_{b}({\mathbf r}) + \frac{\delta}{2}]\hat{\psi}_\uparrow \nonumber \\
    &  +(g_1 \hat{a}e^{ik_Ly} + g_2 \hat{a}^\dagger e^{-ik_Ly})\cos(k_Lx)  \hat{\psi}_\downarrow \nonumber \\
    & + (g_{\uparrow\uparrow}\hat{\psi}_\uparrow^\dag \hat{\psi}_\uparrow+ g_{\uparrow\downarrow}\hat{\psi}_\downarrow^\dag \hat{\psi}_\downarrow )\hat{\psi}_\uparrow, \nonumber \\
    %%%%%%%%%%%%%%%%%%%%%%%%%%%%%%%%%%%%%%%%%%%%%%%%%%%%%%%%%%%%%%%%%
    i\dot{\hat{\psi}}_\downarrow & = [-U_0\hat{a}^\dagger \hat{a}\cos^2(k_Lx) + V_{b}({\mathbf r}) - \frac{\delta}{2}]\hat{\psi}_\downarrow \nonumber \\
    &+(g_1 \hat{a}^\dagger e^{-ik_Ly} + g_2 \hat{a} e^{ik_Ly})\cos(k_Lx)  \hat{\psi}_\uparrow  \nonumber \\
    &+ (g_{\downarrow\downarrow}\hat{\psi}_\downarrow^\dag \hat{\psi}_\downarrow+ g_{\uparrow\downarrow}\hat{\psi}_\uparrow^\dag \hat{\psi}_\uparrow )\hat{\psi}_\downarrow,
     \label{smdyn}
\end{align}
where $\tilde{\Delta}_c =(\Delta_c - U_0\hat{\cal B})$ is the effective cavity detuning with
$\hat{\cal B}=\int d {\mathbf r}\cos^2(k_L x)\sum_{\sigma}\hat{\psi}_\sigma^\dag({\mathbf r})\hat{\psi}_\sigma ({\mathbf r})$ and ${\hat{\Xi}_1}  = \int d {\mathbf r} \cos(k_Lx) e^{ik_Ly} \hat{\psi}_\uparrow^\dag({\mathbf r})\hat{\psi}_\downarrow({\mathbf r})$ and ${\hat{\Xi}_2}  = \int d {\mathbf r} \cos(k_Lx) e^{-ik_Ly}\hat{\psi}_\uparrow^\dag({\mathbf r})\hat{\psi}_\downarrow({\mathbf r})$ are introduced order parameters that determine the configuration of SS phase for superradiance.

Since the cavity field quickly reaches a steady state which is much faster than the external atomic motion in the far dispersive regime with $|\tilde{\Delta}_c/g_{1,2}|\gg 1$~\cite{RevModPhys.85.553,Mivehvar:2021lpo}, the steady-state photon number $\alpha_{}=\langle\hat{a}_{}\rangle$ for cavity field can be written as
\begin{align}\label{smcavity}
    \hat{a} &= \frac{g_1 {\hat{\Xi}_1^\dagger+g_2 {\hat{\Xi}_2}}}{-\tilde{\Delta}_c +
    i\kappa}.
\end{align}

In order to understand the property of the system more clearly from the physical image, we derive the cavity-mediated long-range interactions of atomic fields by integrating out the cavity with the steady-state solution Eq.~(\ref{smcavity}), which yields 
\begin{align}\label{longrange}
 \hat{\mathcal{H}}_{2}&=\frac{1}{2}\int d\mathbf{r}d\mathbf{r^\prime} U_{\rm eff}(\mathbf{r},\mathbf{r^\prime}) \{V_1\hat{S}_-(\mathbf{r})\hat{S}_+(\mathbf{r^\prime})  \nonumber \\
 & + V_2\hat{S}_+(\mathbf{r})\hat{S}_-(\mathbf{r^\prime}) + V_3[\hat{S}_+(\mathbf{r})\hat{S}_+(\mathbf{r^\prime})+ {\rm H.c.}] \},
\end{align}
where $V_{j=1,2} =-\frac{\tilde{\Delta}_cg_j^2}{\tilde{\Delta}_c^2 + \kappa^2}, V_3= -\frac{\tilde{\Delta}_cg_2g_1}{\tilde{\Delta}_c^2 + \kappa^2}$ are the tunable strengths of the cavity-mediated atom-atom interaction, $U_{\rm eff}(\mathbf{r},\mathbf{r^\prime})=2\cos(k_Lx)\cos(k_Lx^\prime)e^{ik_L(y-y^\prime)}$ is the two-body long-range potential and $S_-^\dagger(\mathbf{r})=S_+(\mathbf{r})=\hat{\psi}_\uparrow^\dagger(\mathbf{r})\hat{\psi}_\downarrow(\mathbf{r})$ is the spin operator. Here the first two terms of $\mathcal{H}_{2}$ is the long-range spin-exchange interaction between spin-$\uparrow$ and spin-$\downarrow$ atoms conserving the atomic number in the individual spin state. The third term represents the long-range two-axis twisting interaction via a superradiant photon-exchange process and does not conserve the atom number in the individual spin state.

Furthermore, we map the many-body Hamiltonian to two-component Tavis-Cummings model (TCM) in the single recoil scattering limit~\cite{baumann2010dicke} since the other higher order modes induced by the Raman process of cavity two-body collisional interaction can be neglected. We consider a homogeneous BEC initially prepared in $|\uparrow\rangle$ state and the atomic field operator $\hat\psi_{\sigma}$ ($\sigma=\uparrow, \downarrow$) can be expanded as
\begin{align}\label{exmod}
    \hat\psi_{\uparrow}&=\sqrt{\frac{1}{{{V}}}}\hat b_{\uparrow,0}, \nonumber \\
    \hat \psi_{\downarrow}&=\sqrt{\frac{2}{V}}\cos(k_Lx)\left(e^{-ik_Ly}\hat b_{\downarrow,1}+ e^{ik_Ly}\hat b_{\downarrow,2}\right),
\end{align}
where $\hat b_{\downarrow,1}$ ($\hat b_{\downarrow,2}$) represents the bosonic operator corresponding to the relevant atomic momentum modes $|k_{x},k_{y}\rangle=|\pm k_L, k_L\rangle$ ($|\pm k_L, -k_L\rangle$) for one of spin-$\downarrow$ atoms, $\hat b_{\uparrow,0}$ represents the bosonic operator corresponding to the relevant atomic momentum mode $|0, 0\rangle$ for spin-$\uparrow$ atoms,  with $V$ being the volume of condensates and the total atom number is $N=\hat b_{\uparrow,0}^\dag\hat b_{\uparrow,0} + \hat b_{\downarrow,1}^\dag\hat b_{\downarrow,1}+ \hat b_{\downarrow,2}^\dag\hat b_{\downarrow,2}$. Excluding external trapping potential and collisional interaction,  with expansion, the emerged two-component TCM Hamiltonian originating from the many-body Hamiltonian (\ref{smmany}) is given by ($\hbar =1$)
\begin{align}\label{SMTC}
\hat{\mathcal{H}_1}& =\! \tilde{\Delta}_c \hat{a}^{\dagger} \hat{a}  +\! \omega_0\hat{J}_z +\! \frac{g_1}{\sqrt{2}}(\hat{a}\hat{J}_-^{(1)} \!+\! \frac{g_2}{g_1}\hat{a}^\dagger\hat{J}_-^{(2)} \!+\! {\rm H.c.} ),
\end{align}
where $\hat{J}_{-}^{(1)} = \hat b_{\uparrow,0}^\dag\hat b_{\downarrow,1}$, $\hat{J}_{-}^{(2)} = \hat b_{\uparrow,0}^\dag\hat b_{\downarrow,2}$ and $\hat{J}_{z} = ( \hat b_{\downarrow,1}^\dag\hat b_{\downarrow,1} +\hat b_{\downarrow,2}^\dag\hat b_{\downarrow,2}- \hat b_{\uparrow,0}^\dag\hat b_{\uparrow,0})/2$ are the collective spin operators. And $\omega_0= 2E_L/\hbar-\delta$ is the effective detuning of atomic field.

The $\hat{a}\hat{J}_{-}^{(1)}$ (non-rotating wave coupling terms) in Hamiltonian (\ref{SMTC}) does not conserve the number of total excitations~\cite{PhysRevResearch.5.013002}. Interestingly, we find that the two-component TCM Hamiltonian possesses a $U(1)$ symmetry characterized by the action of the operator
\begin{align}
{\cal R}_{\theta} =\exp[i\theta(\hat{a}^\dag\hat{a}-\hat{b}_{\downarrow,1}^\dagger\hat{b}_{\downarrow,1}+\hat{b}_{\downarrow,2}^\dagger\hat{b}_{\downarrow,2})],
\end{align}
which yields 
\begin{align}
 {\cal R}_{\theta}^\dag(\hat{a},\hat{J}^{(1)}_-
    ,\hat{J}^{(2)}_-){\cal R}_{\theta} =(\hat{a} e^{i\theta},\hat{J}^{(1)}_-e^{-i\theta},\hat{J}^{(2)}_-e^{i\theta}),
\end{align}
for arbitrary rotational angle $\theta$.

To gain more insight, we insert the expansion Eq.~(\ref{exmod}) back into the cavity-mediated collective interaction of Eq.~(\ref{longrange}). As a result, the spin-momentum mixing interactions in terms of spin and momentum modes are described by
\begin{align}\label{spinmixing_SM}
    \hat{\mathcal{H}}_{3}&=\sum_{j=1,2}\frac{V_j}{2}\hat{b}_{\uparrow,0}^\dagger\hat{b}_{\downarrow,j}^\dagger\hat{b}_{\uparrow,0}\hat{b}_{\downarrow,j} +\!\frac{V_3}{2}(\hat{b}_{\uparrow,0}^\dagger\hat{b}_{\uparrow,0}^\dagger\hat{b}_{\downarrow,1}\hat{b}_{\downarrow,2}+\!{\rm H.c.}).
\end{align}
The first term denotes collective momentum-exchange interactions~\cite{luo2023cavity}, referred to as one-axis twisting. Notably, the second term originating from two-axis twisting provides a new source for the deterministic generation of entangled momentum-correlated pairs from zero-momentum condensates. This controllable spin-momentum-mixing interactions is similar to spin-mixing in spin-1 BEC under the single-mode approximation~\cite{PhysRevLett.81.5257}. Significantly, the magnitude of $N{V}_3$ typically in the order of tens of kilohertz ($20 E_L$), vastly exceeds the spin-mixing rate (few ten hertz) achievable in experimental engineering deterministic entanglement in $^{87}$Rb condensates~\cite{luo2017deterministic}. 

For fixing $g_1=g_2$, we can further derive an effective XX-Heisenberg Hamiltonian in well-defined orbital angular momentum operators 
\begin{align}\label{smorbital}
  \hat{\mathcal{H}}_{\rm eff}&=\frac{1}{4}V_3\hat{L}_+\hat{L}_-=\frac{1}{4}V_3(\hat{L}^2 -\hat{L}_z^2+ \hat{L}_z),
\end{align}
where the raising and lowering operators $\hat{L}_+=\hat{L}_-^\dagger=\sqrt{2}(\hat{b}^\dagger_{\downarrow,2}\hat{b}_{\uparrow,0}+\hat{b}^\dagger_{\uparrow,0}\hat{b}_{\downarrow,1})$  obey the angular momentum commutation relations: $[\hat{L}_+,\hat{L}_-]=2\hat{L}_z$, $[\hat{L}_z,\hat{L}_\pm]=\pm \hat{L}_\pm$ with $\hat{L}_z=\hat{b}_{\downarrow,1}^\dagger\hat{b}_{\downarrow,1}-\hat{b}_{\downarrow,2}^\dagger\hat{b}_{\downarrow,2}$. The Hamiltonian (\ref{smorbital}) is deeply entangled and enables generation of novel quantum states that are highly-correlated in their spin and motional degrees of freedom simultaneously. 

\section{Superradiant phase transition}
\begin{figure}
    \includegraphics[width=0.85\columnwidth]{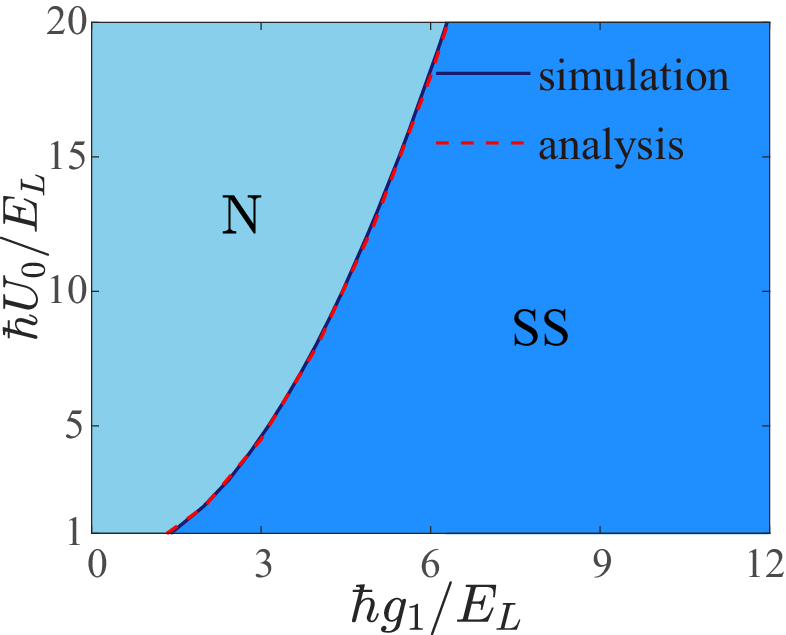}%
    \caption{(Color online). Ground-state phase diagram on $g_1$-$U_0$ parameter plane. The solid (dashed) line denotes the numerical (analytical) result of superradiant phase boundary. }\label{app-phase}
\end{figure}

In this section, we outline the derivation of the quantum phase transition for two-component TCM. In the thermodynamic limit with $N\rightarrow \infty$, we can apply the Holstein-Primarkoff transformation
\begin{align}
    \hat{J}_-^{(1)}&=\sqrt{N-\hat{b}^\dagger_{1}\hat{b}_{1}-\hat{b}^\dagger_{2}\hat{b}_{2}}\; \hat{b}_1\nonumber\\\nonumber
    \hat{J}_-^{(2)}&=\sqrt{N-\hat{b}^\dagger_{1}\hat{b}_{1}-\hat{b}^\dagger_{2}\hat{b}_{2}} \;\hat{b}_2\nonumber\\
    \hat{J}_z&=\hat{b}^\dagger_{1}\hat{b}_{1}+\hat{b}^\dagger_{2}\hat{b}_{2}-\frac{N}{2}\label{HP},
\end{align}
where the bosonic operators $\hat{b}_i$ and $\hat{b}_{i^\prime}^\dagger (i, i^\prime = 1, 2)$ satisfy the commutation relation $[\hat{b}_i,\hat{b}_{i^\prime}^\dagger]=\delta_{ii^\prime},~[\hat{b}_i,\hat{b}_{i^\prime}]=0$, and $[\hat{b}_i^\dagger,\hat{b}_{i^\prime}^\dagger]=0$ and the subscript $\downarrow$ has been neglected for shorthand notation. 
In the weak excited approximation $\sqrt{N-\hat{b}^\dagger_{1}\hat{b}_{1}-\hat{b}^\dagger_{2}\hat{b}_{2}}\approx \sqrt{N}$, the Hamiltonian (\ref{SMTC}) can be transformed into
\begin{align}\label{weakH}
    \hat{\mathcal{H}}^{(1)}/\hbar& =\omega_1 \hat{a}^{\dagger} \hat{a}  + \omega_0(\hat{b}_1^\dagger\hat{b}_1 + \hat{b}_2^\dagger\hat{b}_2)\nonumber\\
    & + \frac{1}{\sqrt{2}}\left[ \lambda_1\hat{a}\hat{b}_1 + \lambda_2{\hat{a}^\dagger}\hat{b}_2 + {\rm H.c.} \right], 
\end{align}
where $\lambda_{1,2}=g_{1,2}\sqrt{N}$ and $\omega_1=\tilde{\Delta}_c$ are introduced for shorthand notation. 

\begin{figure}
    \includegraphics[width=0.99\columnwidth]{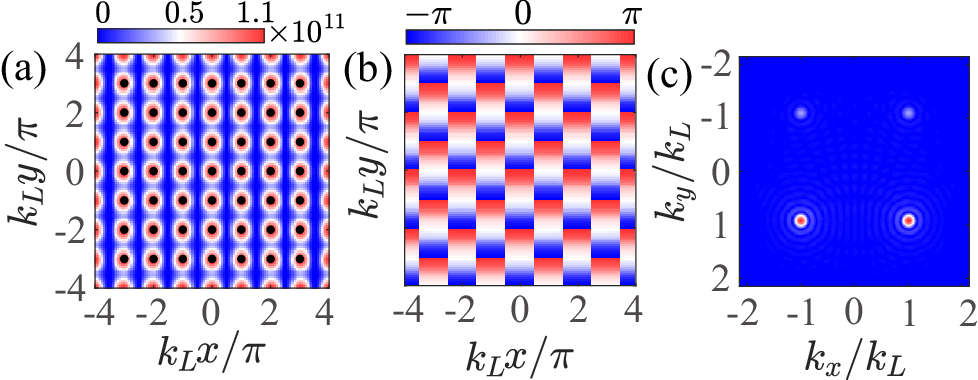}
    \caption{(Color online). The condensates wave function for SS phase with $(g_1, g_2) = (4.2, 7)E_L/\hbar$. Column 1-3 show the density $\rho_\downarrow$ (units of ${\rm cm}^{-2}$), relative phase $\Delta \phi$, and momentum distribution of the $|\downarrow\rangle$ state with the peak areas representing high atomic density. The black dots in (a) denotes the position of sites satisfying $\cos(k_L x)\cos(k_L y) = \pm 1$.}\label{app-rho}
\end{figure}

%\begin{figure}
%    \includegraphics[width=0.95\columnwidth]{square_mmp.eps}
%    \caption{(Color online). The atomic momentum-space distribution of $\psi_\downarrow(\bf{r})$ for the self-organized PW phase with $(g_1, g_2) = (0, 7)E_L/\hbar$ and the SS phase with $(g_1, g_2) = (7, 7)E_L/\hbar$, respectively. The blue-to-red gradient shading indicates the occupation probability $|\psi_\downarrow|$.}\label{app-mmp}
%\end{figure}

Quantitatively, Hamiltonian (\ref{weakH}) can be diagonalized by the introduction of a position-momentum representation for seeking the instability point
\begin{align}
   X=\sqrt{\frac{\hbar}{2m\omega_1}}(\hat{a}^{\dagger}+\hat{a})&;\; P_x = i\sqrt{\frac{m\hbar\omega_1}{2}}(\hat{a}^{\dagger}-\hat{a}) \nonumber \\
    Y=\sqrt{\frac{\hbar}{2m\omega_0}}(\hat{b}^{\dagger}_1+\hat{b}_1)&;\; P_y = i\sqrt{\frac{m\hbar\omega_0}{2}}(\hat{b}^{\dagger}_1-\hat{b}_1)\nonumber \\
     Z=\sqrt{\frac{\hbar}{2m\omega_0}}(\hat{b}^{\dagger}_2+\hat{b}_2)&;\; P_z = i\sqrt{\frac{m\hbar\omega_0}{2}}(\hat{b}^{\dagger}_2-\hat{b}_2)
\end{align}
which yields
\begin{align}
    \hat{a} &= \frac{1}{\sqrt{2}} (\sqrt{\frac{m\omega_1}{\hbar}}X + \frac{i}{\sqrt{m\hbar\omega_1}}P_x) \nonumber \\
    \hat{b}_1 &= \frac{1}{\sqrt{2}} (\sqrt{\frac{m\omega_0}{\hbar}}Y + \frac{i}{\sqrt{m\hbar\omega_0}}P_y) \nonumber \\
    \hat{b}_2 &= \frac{1}{\sqrt{2}} (\sqrt{\frac{m\omega_0}{\hbar}}Z + \frac{i}{\sqrt{m\hbar\omega_0}}P_z).
\end{align} 
As a result, the Hamiltonian in terms of the above operators is given by
\begin{align}
    \hat{\mathcal{H}}^{(1)}&= \frac{1}{2}m\omega_1^2X^2+ \frac{P_x^2}{2m}+\frac{1}{2}m\omega_0^2(Y^2+Z^2)+ \frac{P_y^2+P_z^2}{2m}\nonumber \\
    &+\frac{\lambda_1}{\sqrt{2}}(m\sqrt{\omega_1\omega_0}XY-\frac{P_x P_y}{m\sqrt{\omega_1\omega_0}})\nonumber \\
    &+\frac{\lambda_2}{\sqrt{2}}(m\sqrt{\omega_1\omega_0}XZ+\frac{P_x P_z}{m\sqrt{\omega_1\omega_0}})\nonumber \\
    &=\frac{1}{2}m(\omega_1^2 - \frac{\omega_1(\lambda_1^2+\lambda_2^2)}{2\omega_0})X^2\nonumber\\
    &+\frac{1}{2}m\omega_0^2(Y+\lambda_1\sqrt{\frac{\omega_1}{\omega_0}}X)^2
    +\frac{1}{2}m\omega_0^2(Z+\lambda_2\sqrt{\frac{\omega_1}{\omega_0}}X)^2\nonumber\\
    &+\frac{1}{2m}(1-\frac{\lambda_1^2+\lambda_2^2}{2\omega_1\omega_0})P_x^2
    +\frac{1}{2m}(P_y-\frac{\lambda_1}{\sqrt{2\omega_1\omega_0}}P_x)^2\nonumber\\
    &+\frac{1}{2m}(P_z+\frac{\lambda_2}{\sqrt{2\omega_1\omega_0}}P_x)^2.
\end{align}
By diagonalizing the above Hamiltonian, we find that the system will exhibit the instability when the superradiant quantum phase transition occurs, corresponding to the critical thresholds for Raman coupling
\begin{align}\label{cri}
    [g_1^2+g_2^2]_{\rm cr}=2(\tilde{\Delta}_c^2+\kappa^2)^{1/2}\omega_0/N
\end{align}
with the cavity decay $\kappa$ included.

Figure \ref{app-phase} shows the phase diagram of condensates-cavity system on $g_1$-$U_0$ parameter plane with fixing $g_1/g_2=1$. Clearly, the analytic threshold of Eq.~(\ref{cri}) for superradiant quantum phase transition (dashed line) is in high agreement with numerical result (solid line), even ignoring the cavity decay and atom collisions. In Fig. \ref{app-rho}, we plot typical SS phase with density distribution, relative phase distribution, and Fourier transformation of the $|\downarrow\rangle$ state under the different Raman coupling strengths ($g_1/g_2\neq 1$). It is shown that the sites of peak density satisfying $\cos(k_Ly)\cos(k_Lx)=+1(-1)$ with an obviously periodic density modulation of crystalline order is confirmed. Moreover, this setup yields atomic momentum distribution at $(k_x, k_y)=(\pm k_L, \pm k_L)$ and matches the configurations of crystalline square lattice pattern.

\section{Collective excitations for gapless Goldstone mode}
To demonstrate the rigidity of self-ordered superradiant phases, we calculate the collective excitations of cavity-condensates system. One can substitute the Holstein-Primarkoff transformation of Eq.~(\ref{HP}) back into the two-component TCM Hamiltonian (\ref{SMTC}) and displace the bosonic operators $\hat{a}$, $\hat{b}_1$ and $\hat{b}_2$ with respect to their mean values in the following ways~\cite{PhysRevE.67.066203}
\begin{align*}
 \hat{a}^{\dagger} \rightarrow \hat{d}^{\dagger} +\sqrt{\zeta},~~
    \hat{b}^\dagger_{1} \rightarrow \hat{e}^\dagger_{1} -\sqrt{\beta},~~
    \hat{b}^\dagger_{2} \rightarrow \hat{e}^\dagger_2 -\sqrt{\gamma},
\end{align*}
where $\hat{d}$, $\hat{e}_1$ and $\hat{e}_2$ denotes the photonic and atomic quantum fluctuations around its mean-field values with $\langle\hat{a}\rangle=\sqrt{\zeta}$, $\langle\hat{b}_1\rangle= \sqrt{\beta}$ and $\langle\hat{b}_2\rangle= \sqrt{\gamma}$, respectively. 

In the thermodynamic limit by expanding the square root $\sqrt{\xi}\equiv \sqrt{N-\hat{b}^\dagger_{1}\hat{b}_{1}-\hat{b}^\dagger_{2}\hat{b}_{2}}$,  we can safely neglect the higher order terms, which yields
\begin{align}
    \sqrt{\xi}&\approx\sqrt{k} \Big\{ 1-\frac{1}{2k}[\hat{e}^\dagger_1\hat{e}_1 +\hat{e}^\dagger_2\hat{e}_2\nonumber\\\nonumber
     &-\sqrt{\beta}(\hat{e}^\dagger_1+\hat{e}_1)-\sqrt{\gamma}(\hat{e}^\dagger_2+\hat{e}_2)] \\\nonumber
     &-\frac{1}{8k^2}[\sqrt{\beta}(\hat{e}^\dagger_1+\hat{e}_1)+\sqrt{\gamma}(\hat{e}^\dagger_2+\hat{e}_2)]^2 \Big\},
\end{align}
with $k\equiv N-\gamma-\beta$. 
After the displacements and retaining up to second order, we obtain the reduced Hamiltonian
\begin{align}
    \mathcal{H}^{(2)}/\hbar&=\omega_1[\hat{d}^{\dagger}\hat{d}+\sqrt{\zeta}(\hat{d}^{\dagger}+\hat{d})+\zeta] +\omega_0(\beta+\gamma)\nonumber\\
    &+\omega_0[\hat{e}^\dagger_1\hat{e}_1+\hat{e}^\dagger_2\hat{e}_2- \sqrt{\beta}(\hat{e}^\dagger_1+\hat{e}_1)-\sqrt{\gamma}(\hat{e}^\dagger_2+\hat{e}_2)]\nonumber\\
    &+\frac{1}{\sqrt{2N}}[\lambda_1(\hat{d}+\sqrt{\zeta})\sqrt{\xi}(\hat{e}_1-\sqrt{\beta})\nonumber\\
    &+\lambda_2(\hat{d}^\dagger+\sqrt{\zeta})\sqrt{\xi}(\hat{e}_2-\sqrt{\gamma})+{\rm H.c.}].
\end{align}

For minimizing the ground state energy, the linear terms in the bosonic operators in Hamiltonian will be zero. As a result, the steady-state solutions are given by
\begin{align}\label{mv}
    \sqrt{\beta}&=\frac{\lambda_1}{\sqrt{\lambda_1^2+\lambda_2^2}}\sqrt{(1-\mu_0)N/2}, \nonumber \\
    \sqrt{\gamma}&=\frac{\lambda_2}{\sqrt{\lambda_1^2+\lambda_2^2}}\sqrt{(1-\mu_0)N/2}, \nonumber \\
    \sqrt{\zeta}&=\sqrt{\frac{N}{4}(1-\mu_0^2)}\sqrt{\frac{\omega_0}{\omega \mu_0}}.
\end{align}
Inserting the solution back into the expanded Hamiltonian, the quadratic Hamiltonian describing the collective excitations takes the form as
\begin{align}\label{qH}
    \hat{\mathcal{H}}^{(2)}/\hbar&= \omega_1\hat{d}^{\dagger}\hat{d}+\omega_0\frac{1+\mu_0}{2\mu_0}(\hat{e}^\dagger_1\hat{e}_1+\hat{e}^\dagger_2\hat{e}_2) \nonumber \\
    & -\frac{\omega_0 N}{4\mu_0}(1-\mu_0^2)+\frac{\sqrt{1+\mu_0}}{2}\lambda_1{(\hat{d}\hat{e}_1+\hat{e}^\dagger_1\hat{d}^{\dagger})} \nonumber \\
    &+\frac{\sqrt{1+\mu_0}}{2}\lambda_2(\hat{d}^{\dagger}\hat{e}_2+\hat{e}^\dagger_2\hat{d}) \nonumber \\
    &+\frac{(3+\mu_0)(1-\mu_0)}{16\omega_1(1+\mu_0)}[\lambda_1(\hat{e}^\dagger_1+\hat{e}_1)+\lambda_2(\hat{e}^\dagger_2+\hat{e}_2)]^2 \nonumber \\
    &-\frac{1-\mu_0}{4\sqrt{1+\mu_0}}(\hat{d}^{\dagger}+\hat{d})[\lambda_1(\hat{e}^\dagger_1+\hat{e}_1)+\lambda_2(\hat{e}^\dagger_2+\hat{e}_2)],
\end{align}
where $\mu_0 = (1/\mu_1+1/\mu_2)^{-1}$, {$\mu_1 =2\omega_1\omega_0/\lambda_1^2$ and $\mu_2 =2\omega_1\omega_0/\lambda_2^2$}. 
{For simplicity}, the Hamiltonian of Eq.~(\ref{qH}) can be rewritten as
\begin{align}
    \hat{\mathcal{H}}^{(2)}/\hbar&= \omega_1\hat{d}^{\dagger}\hat{d}+\omega_2(\hat{e}^\dagger_1\hat{e}_1+\hat{e}^\dagger_2\hat{e}_2)-{\frac{\omega_0 N}{4\mu_0}}(1-\mu_0^2) \nonumber \\
    +{P_1}&(\hat{d}^{\dagger}+\hat{d})(\hat{e}^\dagger_1+\hat{e}_1)+{P_2}(\hat{d}^{\dagger}+\hat{d})(\hat{e}^\dagger_2+\hat{e}_2) \nonumber \\
    +\Omega_3&(\hat{e}^\dagger_1+\hat{e}_1)^2+\Omega_4(\hat{e}^\dagger_2+\hat{e}_2)^2+2\Omega_x(\hat{e}^\dagger_1+\hat{e}_1) \nonumber \\
    \times(\hat{e}&^\dagger_2+\hat{e}_2)+Q_1({\hat{d}\hat{e}_1+\hat{e}^\dagger_1\hat{d}^{\dagger}})+Q_2(\hat{d}^{\dagger}\hat{e}_2+\hat{e}^\dagger_2\hat{d})
\end{align}
where
\begin{align}
    \omega_2&\equiv\omega_0\frac{1+\mu_0}{2\mu_0}, \nonumber \\
    P_1&\equiv -\frac{1-\mu_0}{4\sqrt{1+\mu_0}}\lambda_1, \nonumber \\
    P_2&\equiv-\frac{1-\mu_0}{4\sqrt{1+\mu_0}}\lambda_2, \nonumber \\
    \Omega_3&\equiv\frac{(3+\mu_0)(1-\mu_0)}{16\omega_1(1+\mu_0)}\lambda_1^2, \nonumber \\
    \Omega_4&\equiv\frac{(3+\mu_0)(1-\mu_0)}{16\omega_1(1+\mu_0)}\lambda_2^2, \nonumber \\
    \Omega_{x}&\equiv\frac{(3+\mu_0)(1-\mu_0)}{16\omega_1(1+\mu_0)}\lambda_1\lambda_2, \nonumber \\
    Q_1&\equiv\frac{\sqrt{1+\mu_0}}{2}\lambda_1, \nonumber \\
    Q_2&\equiv\frac{\sqrt{1+\mu_0}}{2}\lambda_2. \nonumber
\end{align}
are introduced for shorthand notation. Notably, $\mu_0=1$ is equivalent to Eq.~(\ref{cri}) and can be written as
\begin{align}
    \mu_0&=\frac{2\omega_0\omega_1}{(g_1^2+g_2^2)N}=\frac{[g_1^2+g_2^2]_{\rm cr}}{g_1^2+g_2^2},
\end{align}
which represents the threshold of the cavity-condensates system's superradiant transition and this approach calculating the collective excitation spectra in superradiant phase remains valid when $\mu_0< 1$. For PW phase when $g_1/g_2\gg 1$($g_1/g_2\ll 1$), $\mu_0$ degrades to $\mu_1$($\mu_2$) satisfying $\mu_1<1$($\mu_2<1$) and the approach calculating the spectra of TCM equivalently still remains valid. With respect to $\mu_0>1$ for normal phase, Eq.~(\ref{mv}) has no physical meaning and the solution $\zeta=\beta=\gamma=0$ is applied for calculating. 
\begin{widetext}

For superradiance phase, the Heisenberg equations of motion of the quantum fluctuations in the photonic and atomic field operators become{
\begin{align}
    i\frac{\partial}{\partial t}\hat{d} &= \omega_1 \hat{d} + P_1\hat{e}_1 + (P_2+Q_2)\hat{e}_2+(P_1+Q_1)\hat{e}^\dagger_1+P_2\hat{e}^\dagger_2\nonumber\\\notag
    i\frac{\partial}{\partial t}\hat{e}_1 &= P_1\hat{d}+(\omega_2+2\Omega_3)\hat{e}_1 + 2\Omega_{x}\hat{e}_2+(P_1+Q_1)\hat{d}^{\dagger}+2\Omega_3\hat{e}^\dagger_1 + 2\Omega_{x}\hat{e}^\dagger_2\\\notag
    i\frac{\partial}{\partial t}\hat{e}_2 &= (P_2+Q_2)\hat{d}+2\Omega_{x}\hat{e}_1 + (\omega_2+2\Omega_4)\hat{e}_2+P_2\hat{d}^{\dagger}+2\Omega_{x}\hat{e}^\dagger_1 + 2\Omega_{4}\hat{e}^\dagger_2\\\notag
    i\frac{\partial}{\partial t}\hat{d}^{\dagger} &= -\omega_1^* \hat{d}^{\dagger}- P_1^*\hat{e}^\dagger_1 - (P_2^*+Q_2^*)\hat{e}^\dagger_2-(P_1^*+Q_1^*)\hat{e}_1-P_2^*\hat{e}_2\\\notag
    i\frac{\partial}{\partial t}\hat{e}^\dagger_1 &= -P_1^*\hat{d}^{\dagger}-(\omega_2^*+2\Omega_3^*)\hat{e}^\dagger_1 - 2\Omega_{x}^*\hat{e}^\dagger_2-(P_1^*+Q_1^*)\hat{d}-2\Omega_3^*\hat{e}_1 - 2\Omega_{x}^*\hat{e}_2\\
    i\frac{\partial}{\partial t}\hat{e}^\dagger_2 &= -(P_2^*+Q_2^*)\hat{d}^{\dagger}-2\Omega_{x}^*\hat{e}^\dagger_1 - (\omega_2^*+2\Omega_4^*)\hat{e}^\dagger_2-P_2^*\hat{d}-2\Omega_{x}^*\hat{e}_1 - 2\Omega_{4}^*\hat{e}_2.
\end{align}
}Furthermore, we recast these equations in the form of Hopfield-bogoliubov matrix for superradiance phase{
\begin{align}\label{HP2}
    \left(
        \begin{array}{cccccc}
            \omega_1&P_1&P_2+Q_2&0&P_1+Q_1&P_2\\
            P_1&\omega_2+2\Omega_3&2\Omega_{x}&P_1+Q_1&2\Omega_3&2\Omega_{x}\\
            P_2+Q_2&2\Omega_{x}&\omega_2+2\Omega_4&P_2&2\Omega_{x}&2\Omega_4\\
            0&-P_1^*-Q_1^*&-P_2^*&-\omega_1^*&-P_1^*&-P_2^*-Q_2^*\\
            -P_1^*-Q_1^*&-2\Omega_3^*&-2\Omega_{x}^*&-P_1^*&-\omega_2^*-2\Omega_3^*&-2\Omega_{x}^*\\
            -P_2^*&-2\Omega_{x}^*&-2\Omega_4^*&-P_2^*-Q_2^*&-2\Omega_{x}^*&-\omega_2^*-2\Omega_4^*
        \end{array}
    \right)
    \left(
        \begin{array}{c}
            \hat{d}\\
            \hat{e}_1\\
            \hat{e}_2\\
            \hat{d}^{\dagger}\\
            \hat{e}^\dagger_1\\
            \hat{e}^\dagger_2
        \end{array}
    \right)
    =
    \epsilon
    \left(
        \begin{array}{c}
            \hat{d}\\
            \hat{e}_1\\
            \hat{e}_2\\
            \hat{d}^{\dagger}\\
            \hat{e}^\dagger_1\\
            \hat{e}^\dagger_2
        \end{array}
    \right).
\end{align}

With respects to normal phase Hamiltonian, the another solution $\zeta=\beta=\gamma=0$ of linear equations corresponds to the reduced weak excited Hamiltonian (\ref{weakH})}. Analogously, the Heisenberg equations of motion for normal phase are given by{
\begin{align}
    i\frac{\partial}{\partial t}\hat{d} &= \omega_1\hat{d} + \tilde{\lambda}_1\hat{e}^\dagger_1 + \tilde{\lambda}_{2}\hat{e}_2\nonumber\\
    i\frac{\partial}{\partial t}\hat{e}_1 &= \omega_0\hat{e}_1 + \tilde{\lambda}_{1}\hat{d}^{\dagger}\nonumber\\
    i\frac{\partial}{\partial t}\hat{e}_2 &= \omega_0\hat{e}_2 +\tilde{\lambda}_{2}\hat{d}\nonumber\\
    i\frac{\partial}{\partial t}\hat{d}^{\dagger} &= -\omega_1^*\hat{d}^{\dagger} - \tilde{\lambda}_{1}\hat{e}_1 - \tilde{\lambda}_{2}\hat{e}_2^\dagger\nonumber\\
    i\frac{\partial}{\partial t}\hat{e}^\dagger_1 &= -\omega_0^*\hat{e}^\dagger_1 - \tilde{\lambda}_{1}\hat{d}\nonumber\\
    i\frac{\partial}{\partial t}\hat{e}^\dagger_2 &= -\omega_0^*\hat{e}^\dagger_2 -\tilde{\lambda}_{2}\hat{d}^\dagger
\end{align}
}with $\tilde{\lambda}_{1,2}=\lambda_{1,2}/\sqrt{2}$. 
With the forms of the Hopfield-bogoliubov matrix, we can recast these equations 
\begin{align}\label{HP1}
    \left({
        \begin{array}{cccccc}
            \omega_1&0&\tilde{\lambda}_{2}&0&\tilde{\lambda}_{1}&0\\
            0&\omega_0&0&\tilde{\lambda}_{1}&0&0\\
            \tilde{\lambda}_{2}&0&\omega_0&0&0&0\\
            0&-\tilde{\lambda}_{1}&0&-\omega_1^*&0&-\tilde{\lambda}_{2}\\
            -\tilde{\lambda}_{1}&0&0&0&-\omega_0^*&0\\
            0&0&0&-\tilde{\lambda}_{2}&0&-\omega_0^*
        \end{array}}
    \right)
    \left(
        \begin{array}{c}
            \hat{d}\\
            \hat{e}_1\\
            \hat{e}_2\\
            \hat{d}^{\dagger}\\
            \hat{e}^\dagger_1\\
            \hat{e}^\dagger_2
        \end{array}
    \right)
    =
    \epsilon
    \left(
        \begin{array}{c}
            \hat{d}\\
            \hat{e}_1\\
            \hat{e}_2\\
            \hat{d}^{\dagger}\\
            \hat{e}^\dagger_1\\
            \hat{e}^\dagger_2
        \end{array}
    \right).
\end{align}
Thus the collective excitation spectra of our model can be conveniently calculated by numerically diagonalizing the Hopfield-bogoliubov matrix with Eq.~(\ref{HP2}) and Eq.~(\ref{HP1}). 
\end{widetext}

{\section{Effective potential for superradiant phase}}
\begin{figure}
    \includegraphics[width=0.95\columnwidth]{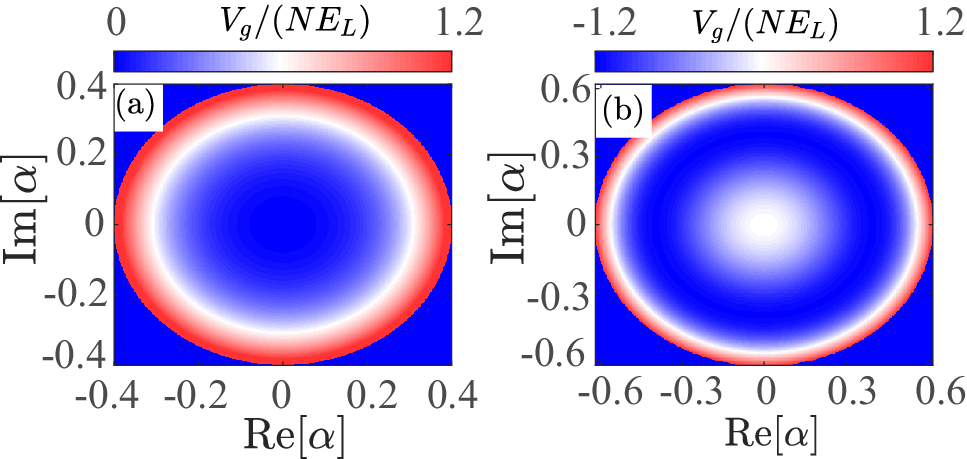}
    \caption{(Color online) The effective potential $V_g$ in normal phase and superradiant phase as {a} function of $\alpha$ (a) at $\hbar g_1/E_L=\hbar g_2/E_L=4$ and (b) at $\hbar g_1/E_L=\hbar g_2/E_L=7.5$, respectively. Here $\hbar U_0/E_L=10$, $N=10^5$, $\hbar\omega_0/E_L=4$ and $\hbar\tilde{\Delta}_c/E_L=5\times 10^5$.\label{effV}}
\end{figure}
For comparison with our above analysis, we perform a stability analysis in a mean-field description by inserting the mean-field ansatz
\begin{align}
\langle\hat{a}\rangle &= \alpha, \nonumber \\
\langle\hat{b}_1\rangle &=\beta_1, \nonumber \\
\langle\hat{b}_2\rangle &=\beta_2, \nonumber \\
\langle\hat{b}_\uparrow\rangle &= \sqrt{N_{\uparrow}-|\beta_1|^2-|\beta_2|^2},
\end{align}
{into the effective reduced Hamiltonian (\ref{SMTC}). After neglecting a constant term, the cavtiy-condensates system is described by the effective potential 
\begin{align}
    V_g&(\alpha, \beta_1, \beta_2)=\tilde{\Delta}_c|\alpha|^2 + \omega_0(|\beta_1|^2+|\beta_2|^2) \nonumber \\
    &+ \frac{1}{\sqrt{2}}\Big(g_1 \alpha \sqrt{N- |\beta_1|^2 - |\beta_2|^2}\beta_1 \nonumber \\
    &+ g_2 \alpha^* \beta_2 \sqrt{N- |\beta_1|^2 - |\beta_2|^2} +{\rm H.c.}\Big).
    \label{potential}
\end{align}

In the far dispersive regime with the limit $\tilde{\Delta}_c/\omega_0\gg 1$, we can adiabatically eliminate the cavity field with $\partial\alpha^*/\partial t=0$ for minimizing the potential
\begin{align*}
    \alpha &= -(g_1 \sqrt{N- |\beta_1|^2 - |\beta_2|^2}\beta_1^* \\
    &+ g_2 \sqrt{N- |\beta_1|^2 - |\beta_2|^2}\beta_2)/\sqrt{2}\tilde{\Delta}_c.
    % \dot{\beta_1}&= \omega_0 \beta_1 + \frac{g_1}{\sqrt{2}}\alpha^* \sqrt{N- |\beta_1|^2 - |\beta_2|^2}\\
    % \dot{\beta_2}&= \omega_0 \beta_2 + \frac{g_2}{\sqrt{2}}\alpha^* \sqrt{N- |\beta_1|^2 - |\beta_2|^2}
\end{align*}

By substituting $\alpha$ in Eq.~(\ref{potential}), the ground state energy in term of $\beta_1,\beta_2$ can be written as
\begin{align*}
    &V_g(\beta_1, \beta_2) = \omega_0\left[(1 - \frac{1}{\mu_1})|\beta_1|^2+\frac{1}{\mu_1 N}|\beta_1|^4\right]\\
    &+\omega_0\big[(1 - \frac{1}{\mu_2})|\beta_2|^2+\frac{1}{\mu_2 N}|\beta_2|^4\big]\\
    &+ \frac{g_1^2+g_2^2}{2\tilde{\Delta}_c}|\beta_1|^2|\beta_2|^2-\frac{g_1 g_2}{2\tilde{\Delta}_c}(N-|\beta_1|^2-|\beta_2|^2)\\
    &\times(\beta_1\beta_2 + \beta_2^*\beta_1^*).
\end{align*}
For more clearly physical image, we assume $\beta_0\equiv\beta_1=\beta_2^*, g_0\equiv g_2=g_1$ and write down a concise expression:
\begin{align}
    V_g(\beta_0)&=2\omega_0[(1-\frac{1}{\mu})|\beta_0|^2+\frac{2}{\mu N}|\beta_0|^4]
\end{align}
with $\mu= \omega_0\tilde{\Delta}_c/(g_0^2N)$, from which the instability point, namely superradiant quantum phase transition point, localizes at $g_{\rm cr} =\sqrt{{\tilde{\Delta}_c\omega_{0}/N}}$ without considering the cavity dissipation.
Notably, for $\mu>1$ the ground state energy of the system in the normal phase emerges a minimum at the origin while for $\mu<1$ the ground state energy of the SS phase has a series of minima satisfying
\begin{align}
    \alpha&=-g_0N\sqrt{\mu(1-\mu)}e^{i\theta}/\tilde{\Delta}_c\\\nonumber
    \beta_0&=\sqrt{N(1-\mu)/2}e^{-i\theta}
\end{align}
as shown in Fig.~\ref{effV}, which demonstrates that the system spontaneously breaks $U(1)$ symmetry along with the superradiant quantum phase transition. 
}

%\bibliographystyle{apsrev4-1-y}

%\bibliography{cavity_ss}


\end{document}